\documentclass[a4paper,notitlepage, 12pt]{article}
\usepackage{amsfonts}
\usepackage{amssymb}
\usepackage{amsmath}
\usepackage{bm}
\usepackage{graphicx}
\usepackage{slashed}
\usepackage{color}
\usepackage{graphics}

\newcommand{\beq}{\begin{equation}}
\newcommand{\eeq}{\end{equation}}
\newcommand{\be}{\begin{eqnarray}}
 \newcommand{\ee}{\end{eqnarray}}
\newcommand{\ov } {\over }

\def\one{{\hbox{1\kern-.8mm l}}}

\begin{document}
\begin{titlepage}
\bigskip\begin{flushright}
NORDITA-2009-21\\
\end{flushright}
\begin{center}
\vskip 4cm
\Large{String Mass Shifts}
\end{center}
\vskip 1cm
\begin{center}
\large{Diego Chialva \\ 
       {\it Nordita Institute
   AlbaNova University Centre,
   Roslagstullsbacken 23
   SE-106 91 Stockholm, Sweden} \\
        \tt{chialva@nordita.org}}

\end{center}
\date{}

\pagestyle{plain}
\vskip 3cm
\begin{abstract}

We study
closed string one-loop amplitudes in string
theory, in particular 
the average mass shift for states at given mass
and Neveu-Schwarz charges.
Our analysis is based only on well-defined string amplitudes and the
exploitation of symmetries and unitarity properties of the torus
amplitudes.

We obtain the result $\Delta M^2 = - g_s^2 M^{2+{3-D \ov 2}}$ in $D$
space-time dimensions for the
average closed string mas-shift ($\Delta M^2 = - g_s^2 (M^2-Q^2)^{1+{3-D \ov 4}}$
for states with non-zero Neveu-Schwarz charges $Q$).
An interesting picture of one-loop corrections
for the string in non\! -\! supersymmetric configurations comes out:
the dominant 
interactions responsible for these corrections are of long-range type
(namely gravitational)
and it appears that
perturbations theory is generally reliable on the spectrum of massive
string states.

\end{abstract}

\end{titlepage}
\newpage

\tableofcontents

\setcounter{section}{0}
\section{Introduction}
String theory is a promising candidate for a quantum theory of gravity
and all the fundamental interactions. One of its appealing features is
that a perturbative expansion is possible and
formally well-defined. Nonetheless, in fact very few is know in detail, already
at the level of one-loop string amplitudes. 

Knowledge of one-loop corrections (mass shifts) is of great importance
both for fundamental aspects of the theory and for
applications in the absence of a protection mechanism against
renormalization such as supersymmetry. 
At the fundamental level it can confirm or
question the reliability of the formal expansion in powers of the
string coupling, and better shape the region of parameters (coupling)
for which it is valid. Applications, on the other hand,
are represented for example by 
investigations of the 
black holes/string correspondence for non-supersymmetric
configurations \cite{chialvasize, HorPolSelf, DamVenSelf}.

The study of
one-loop string diagrams 
has so far mainly focused on some aspects: first of all, on the 
imaginary part of the diagrams for particular states
\cite{Wilkinson:1989tb, Mitchell:1988qe, Chialva:2003hg}. Indeed, thanks to
the  
optical theorem 
the imaginary part has an on-shell definition, which makes the
computation, 
at least in principle, straightforward.
 
As for the real part, properties such
as the lack of UV divergences, unitarity, conformal invariance,
\ldots were those generally
investigated \cite{Seiberg:1986ea}. Only a
restricted number of works (see \cite{SundbShift} and references) have
coped with the  
computation of the actual magnitude of the corrections. 
In fact those papers have dealt only with particular states
(such as those with maximal angular momentum), because of various
advantages in the technology needed to discuss them. 

General studies
on average mass shifts or similar statistical information on the whole
range of the string massive spectrum have instead developed some sort
of field theory or even semiclassical approximations of the string 
calculation, but
so far none of these attempts has proven fully reliable. In particular
they have generally assumed the predominance of gravitational
interactions in the self-energy of a string, but without verifying it
on well-defined string amplitudes.

The difficulties in coping with the mass shifts of strings are both
technical and conceptual. On the one side, they are the consequence of
the lack of 
definition of the theory so far, still bound to an on-shell
first-quantized version\footnote{Studies on the definition and
  computation of off-shell amplitudes do exist, see for example
  \cite{Liccardo:1999hk} and references, but the techniques are not
  well-formalized and most of the results are in the field theory
  limit $\alpha' \to 0$ and in the bosonic string theory.}, and
on the other, they are due to the complexity 
in the computation of well-defined on-shell
string amplitudes (see
section \ref{oneloopform} for a brief discussion).

This paper provides a rigorous study
of string mass shifts, and in particular
statistical information concerning the average mass shift for the
whole spectrum of massive
states a a given mass. In fact, we will be able to clarify some
of the statements expressed in the past about string mass shifts.

We will investigate closed string one-loop
amplitudes and establish an algorithm to study them efficiently,  
individuating the imaginary part, the real one, the
different asymptotic contributions and finally computing the mass shift
for large masses. Fully determining the dominant contribution 
will turn out to be quite involved,
but we will provide arguments and evidence for our conclusions.

Our algorithm is well suited for both pure string states
and mixed ones. This is of particular interest, since it allows
studying general average properties of string mass shifts. 
We stress, in any case, that the only ingredients in
the algorithm are well defined string off-shell amplitudes. 

The paper is so subdivided: in Section \ref{oneloopform} we review
the string formulas for one-loop amplitudes and summarize some of
their features which will be important in the following. In section
\ref{reANDim} we re-write the one-loop amplitude in a convenient way
by expanding it\footnote{No approximation is involved in this.} 
as a sum of terms constituted by a coefficient and an integral part.
We then study these respectively in sections \ref{realimagpart} and
\ref{coeffgen}. 
We finally apply our techniques to the computation of 
the average mass shift for states at a certain mass level
$N$ and charges $Q^i$ in section \ref{averagemassshift}.   

Finally, we comment and conclude.

\section{String formula for the one-loop amplitude.} \label{oneloopform}

\begin{figure}[t]
 \begin{center}
\includegraphics[height=4cm]{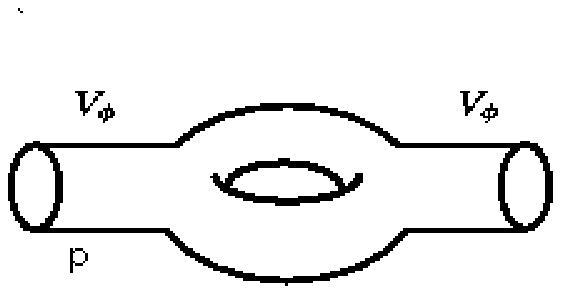}
\caption{{\small \sl{Two-point one-loop amplitude for the state
      represented by the vertex operator $V_\phi$ with four-momentum $p$.}}}  
 \label{torustwopoint}
\end{center}
\end{figure}

Studying the mass shift of fundamental closed strings in perturbative
string theory means
to compute torus amplitudes with the insertions of two vertex
operators representing the string state of interest (see figure
\ref{torustwopoint}). Such calculations are 
difficult to be performed and even defined in string theory for a
series of reasons:
 \begin{itemize}
 \item the form of vertex operators for massive states is complicated
 \item looking for statistical properties means in principle to be
   able to compute one-loop two-points amplitudes for all possible
   string states in an ensemble,  
   but only a few vertex operators are explicitly known
 \item one-loop two-points amplitudes are divergent (due to the presence
     of an imaginary part), therefore issues of analytical
     continuation must be carefully studied. This is difficult within
     String Theory, since in its present status the theory is defined
     only on-shell. 
 \end{itemize}
 
An optimal method for solving (some of) these problems and computing would be
factorization \cite{SundbShift}: starting from a known four-point
amplitude, we can factorize the external legs pairwise and obtain the
squared mass shifts for the intermediate states as the residue of the double
pole for the center of mass energy. In that case we do not need the
detailed knowledge of the form of vertex operators and, as we said
above, the squared momentum
flowing in the loop is now a continuous variable, allowing analytical
continuation. 

Unfortunately this approach has a residual unsolved
problem: in order to identify mass shifts for the
various states we need to know the form of all their couplings with
the external legs of the amplitude\footnote{For
particular states, namely those on the Regge trajectory, which are 
non-degenerate, the method can be implemented, see \cite{SundbShift}.}. 

In this work we will take another approach, considering instead
well-defined one-loop two-point
string amplitudes and extracting from
them the string mass shifts for the 
states of our interest.

Our setup is closed superstring theory\footnote{IIA or
IIB, even though our techniques could be extended to type I theory.}
with
$10-D$ small
compactified dimensions\footnote{We use $D$ for indicating the
  space-time extended 
  dimension, $d=D-1$ for the spatial ones.} (we consider toroidal
compactification, for
simplicity).
The string states taken into account
are extended in the uncompactified dimensions (that is they are
constituted by excitations of the uncompactified string coordinate
operators only), but can possibly wrap or have momentum
charges (Neveu-Schwarz charges) along the compact dimensions.

We work in a {\em time gauge} (\cite{Hwang}), such that the 
{\em on-shell} vertex 
operators have the form\footnote{Here and in the following, our vertex
  operators do not 
  have the usual closed string coupling factor $g_s$ carried by the string
  vertex operators. This is because for clarity we have decided to
  explicitly show all string coupling factors in the formulas for the
  amplitudes.}: 
 \beq \label{genericvertex}
 V_\phi  =  e^{-i p^0 X^0+i \vec p\cdot \vec X}
         V(X^i, \psi^i, S^\alpha) 
 \eeq
where the $S^{\alpha}$ are spin-fields and $i$ runs over the spatial dimensions.
For the bosonic string we would have 
$V_\phi =  e^{-i p^0 X^0+i \vec p\cdot \vec X} V(X^i)$.
 
Furthermore, we choose to work in the center-of-mass reference system,
where
 \beq
 p^\mu = \{p^0,\, \vec 0,\, \{p^j_L, p^j_R\}\}, \qquad 
   j=d+1, d+2, \ldots \,,
 \eeq
where $\vec 0$ gives the momentum spatial components in the extended
dimensions and $j$ runs over the compact ones. 

To simplify the notation, we
  consider initial 
states $\phi$ having only Kaluza-Klein {\em or} winding charges,
not present both at the same time\footnote{In any case, all the
  analysis in this 
work can be repeated for cases where both charges are present at
the same time: it is sufficient to 1) distinguish the right- and
left-moving factors in the amplitude by introducing $N_R$ and $N_L$
respectively instead than $N$, 2) operate two distinct expansions in
formula (\ref{PsExpansion}) with
indexes $s, \tilde s$ and 3) modify accordingly the analysis in
the rest of this work.}. Therefore, being the vertex
operators on-shell,  
 \beq \label{massshellcond}
 (p^0)^2=N+Q^2 \equiv N+\sum_j(p^j_{L(R)})^2\,, \quad 
  |p^j_L|=|p^j_R| \, ,
 \eeq 
where $M_0^2 = N+Q^2$ is the tree-level squared mass.
 
Note that the excitations 
present in $V(X^i, \psi^i, S^\alpha)$ have $i \notin \{d+1, d+2,
\ldots, D\}$, because the states we consider are extended only in the
uncompactified dimensions.
Finally, in our units $\alpha' = 4$. 

The string one-loop on-shell amplitude for a string state represented
by such a vertex operator $V_\phi$ is\footnote{Here and in the following a bar
over a quantity indicates its complex conjugate. Also, the overall
normalization of the one-loop amplitude has been absorbed in the
correlator $\langle V_{\phi}(0) V_\phi(\nu)\rangle_{T^2}$ for
simplicity of notation.}
 \be \label{oneloopfixedRN}
 iT_{T^2} & = & i g_s^2
  \iiint\langle V_{\phi}(0) V_\phi(\nu)\rangle_{T^2} \\ 
  & = & 
  \!ig_s^2
 \int \!\!d^2\!\tau\, d^2\!\nu \,
 {e^{-4\pi (N+Q^2){\text{Im}(\nu)^2 \ov \text{Im}(\tau)}}\ov (\text{Im}(\tau))^{{d+1 \ov 2}}}
   {L(d_c, d, \tau, \bar\tau,\nu, \bar\nu) \ov |\eta(\tau)|^{2(D-2)}}
 \!\!\left|{\theta_{1}(\nu,\tau) \ov \theta'_{1}(0,\tau)} \right|^{4N}
 \!\!\!\!\!\!
 \nonumber \\
 & & ~~~~~~ \times
 \mathcal{P}_\phi(W,\Omega,\partial_\nu \Omega,..,
 \bar\Omega,\partial_{\bar\nu} \bar\Omega,..)\,\nonumber \\
 & & ~~~~~~ \times \mathcal{X}_\phi(\nu, \tau) \tilde{\mathcal{X}}_\phi(\bar\nu, \bar\tau)  
   \nonumber 
 \ee
with 
 \be \label{OmegaW}
 \Omega &= &\partial_\nu^2 \ln(e^{-{\pi\nu^2 \ov \text{Im}(\tau)}}\theta_{1}(\nu,\tau))
  \\
 W & = & {2\pi \ov \text{Im}(\tau)}
 \ee
and it is related to the S-matrix by:
 \beq
  S=\one+i\,T.
 \eeq
In the above formula, $\mathcal{P}_\phi$ is a polynomial of $W\,, \Omega,\, 
 \bar\Omega$ and their higher derivatives
and $\mathcal{X}_\phi (\tilde{\mathcal{X}}_\phi)$ is the (anti)holomorphic
fermionic part of the amplitude\footnote{If we were to study the
  bosonic string theory, the formula to use would be pretty similar,
  with a different $\mathcal{P}_\phi$ and no fermionic
  $\mathcal{X}_\phi (\tilde{\mathcal{X}}_\phi)$.}. 

All the relevant
information 
regarding the state $|\phi\rangle$, contained in the factor
$V(X^i, \psi^i, S^\alpha)$ of the vertex operator (see formula
(\ref{genericvertex})), is stored in the quantities
$\mathcal{P}_\phi, \mathcal{X}_\phi, \tilde{\mathcal{X}}_\phi$.
In particular, the amplitude for mixed string states, represented by a
density matrix such as
 \beq
 \rho = \sum_{\phi} c_\phi |\phi\rangle \langle \phi|,
 \eeq
can be written, using the vertex operator formalism, as the sum over
the one-loop two point functions for the {\em physical} states
$|\phi\rangle$ 
with the relevant coefficients $c_\phi$.  The form of the summed up
 amplitude is therefore analogous to (\ref{oneloopfixedRN}), with
 $\mathcal{P}_\phi, \mathcal{X}_\phi, , \tilde{\mathcal{X}}_\phi$
 replaced by
 $\mathcal{P}_\rho, \mathcal{X}_\rho, \bar{\mathcal{X}}_\rho$  
 depending on $W, \Omega, \bar\Omega$ and higher derivatives, as well as the
 various $\{c_\phi\}$.

If we knew explicitly the form of the vertex operator $V_\phi$, then
we could fully determine
the quantities $\mathcal{P}_\phi, \mathcal{X}_\phi, ,
\tilde{\mathcal{X}}_\phi$
(similarly for mixed states, if we knew the
form of the vertex operators for the states entering the
definition of the density matrix). We will however show
how symmetries and unitarity properties of the torus help in
specifying them up to some elements 
that can be computed directly, even in cases where not all the
relevant vertex operators are known (such as for the string average, 
see section \ref{averagemassshift}). 

In formula (\ref{oneloopfixedRN}), the contribution from the
compactified dimensions is given by 
$L(d_c, d, \tau, \bar\tau, \nu, \bar\nu)$, which, assuming  for simplicity 
compactification on a torus, reads
 \be \label{compactTd}
 L(d_c, d, \tau, \bar\tau) & = & \prod_{i=1}^{d_c-d}\frac{1}{R_{(i)}}
  e^{\sum_{n_i,w_i}2\pi i\tau\sum_i\left({n^i\ov R^i}+{w_iR^i\ov 4}\right)^2-2\pi i\bar\tau\sum_i\left({n^i\ov R^i}-{w_iR^i\ov 4}\right)^2} \\
   & & \times e^{-4\pi i \nu\sum_ip^i_L\left({n^i\ov R^i}+{w_iR^i\ov 4}\right)+4\pi i \bar\nu\sum_i p^i_R\left({n^i\ov R^i}-{w^iR^i\ov 4}\right)},
 \ee
where $d_c$ is the critical (spatial) dimension, that is 9 for the
superstring\footnote{It would be 25 for the
bosonic theory.}.

The coordinate  
$\tau =\tau_1+i\tau_2$ represents 
the torus moduli and is therefore integrated over the complex plane after
dividing out the gauge transformations that preserve the metric (see
\cite{Polchinski}, volume 1, chapter 5), namely, on the torus, modular
transformations. This translates, as we will discuss 
more at length in the following, in a subdivision of the complex
plane in several {\em fundamental
regions} and the integration is restricted to one of these. It is
customary to choose the one defined as
 \beq \label{taureg}
 F=
 \{\tau\,\epsilon\, \mathbb{C}: |\tau| \geq 1, |\tau_1|\leq {1 \ov 2}\}.
 \eeq
On the other hand, 
$\nu= \nu_1 +i\nu_2$ represents the modulus related to the position of
one of the 
vertex operators (the other one is fixed by the conformal Killing
vectors on the torus, see \cite{Polchinski}) and is
integrated over a region
 \beq \label{nureg}
  0 \leq \nu_2 \leq \tau_2, \qquad |\nu_1|\leq {1 \ov 2}.
 \eeq
Finally, a brief note about the singularities of the one-loop
amplitude. 
It is well-known that there
are no UV singularities because they would arise from the region of
integration where $\tau_2 \to 0$, which is eliminated because of
modular invariance (see (\ref{taureg})). 
The remaining singularities
arise all from IR behavior. 

In particular the dangerous limit is
represented by $\nu \to 0$, that is when the two insertion points of
the vertex operators approach each other. The divergence is a tadpole
which signals a modification of the background. Nonetheless in most
supersymmetric backgrounds (such as the one we have here) these IR
divergences actually vanish, while in the non-supersymmetric case they
are understood in terms of
the Fischler-Susskind mechanism
\cite{FischlerSusskind}.  

Having reviewed the general features of one-loop string amplitudes, we
now turn to a detailed study of their real and imaginary
parts\footnote{Recall that due to unitarity, the
  imaginary part of the one-loop S-matrix is related to the sum over
  the final products of the
  squared modulus of the tree-level S-matrix for the decays of the
  initial state: 
   \beq
     T_{T^2, \phi}-\bar T_{T^2, \phi} = i\sum_{\phi', \zeta}
     |T_{\text{tree}, \phi, \phi', \zeta}|^2.
   \eeq}. We will proceed in the following way:
 \begin{itemize}
  \item rewrite the amplitude as a sum over different terms, each
    constituted of a {\em coefficient} part and an {\em integral} over
    the moduli $\tau, \nu$
  \item focus on the integral part distinguishing the real and
    imaginary parts 
  \item focus on the coefficient part
  \item compute the real part in a specific case: the average for
    large fixed mass and fixed charges.
 \end{itemize}

\section{Real and imaginary part of the one-loop amplitude} \label{reANDim}

In this section we are going to obtain the main formulas we will use
in the computation of the 
string one-loop diagrams.

\subsection{Expansion of the one-loop amplitude}
\label{subsec_exponeloop}

We will now analyze formula (\ref{oneloopfixedRN}) and suitably expand its
integrand. In the following we do 
not distinguish between pure and mixed states, since what we will
say is valid in both cases\footnote{The notation will vary
  accordingly: we will neglect the suffixes in $\mathcal{P}_{\phi /\rho},
  \mathcal{X}_{\phi /\rho}$ and simply write $\mathcal{P},
  \mathcal{X}$ instead.}.

To proceed, we need to distinguish the
holomorphic and anti-holomorphic part of the integrand. The
factorization in these two parts
is not exact, due to the presence
of the mixed derivative term $W$, but the integrand can be expanded in
powers of $W$ and 
every term will factorize independently.
Therefore, we write
 \begin{align} \label{PsExpansion}
 \mathcal{P}(W,\Omega,\partial_\nu \Omega,..,
 \bar\Omega,\partial_{\bar\nu} \bar\Omega,..)& =  \\
    =\sum_s& \, W^s(\tau) \mathcal{M}_s (\Omega,\partial_\nu \Omega,\ldots,
   \bar\Omega,\partial_{\bar\nu} \bar\Omega,\ldots) \nonumber \\
   =\sum_{s} & \,
    W^s(\tau) \sum _{u, \tilde u} \Omega^u \bar \Omega^{\tilde u}
    \mathcal{U}_{s, u} (\partial_\nu \Omega,\ldots)
   \bar{\mathcal{U}}_{s, \tilde u}(\partial_{\bar\nu} \bar\Omega,\ldots)\,  . 
   \nonumber
 \end{align}
Using Newton's binomial
 \be
  \Omega^u & = & \left(\partial^2_\nu \log\theta_1 -{2\pi \ov \tau_2}\right)^u 
   \\
    & = & \sum_{f=0}^u \left(\begin{matrix} u \\ f \end{matrix}\right)
    (\partial^2_\nu \log\theta_1)^{u-f}\left({2\pi \ov \tau_2}\right)^f (-1)^f
 \ee
and defining
 \beq \label{defrtilder}
  r \equiv f+s\,, \qquad \tilde r \equiv \tilde f +s
 \eeq
we obtain
 \begin{multline} \label{PinD} 
  \mathcal{P}(W,\Omega,\partial_\nu \Omega,..,
    \bar\Omega,\partial_{\bar\nu} \bar\Omega,..) = \\
  \sum_{s, r, \tilde r} {1 \ov \tau_2^{r+\tilde r-s}} 
    \mathcal{D}^r_s(\partial^2 \log\theta_1, \partial^3 \log\theta_1, \ldots)
    \bar{\mathcal{D}}^{\tilde r}_s(\bar\partial^2 \log\bar\theta_1, \bar\partial^3 \log\bar\theta_1, \ldots)
 \end{multline}
where we have defined
 \begin{multline} \label{defD}
   \mathcal{D}^r_s(\partial^2 \log\theta_1, \partial^3 \log\theta_1,\ldots) = \\
   \sum_u (2 \pi)^{{s \ov 2}} \left(\begin{matrix} u \\ r-s \end{matrix}\right)
  (\partial^2_\nu \log\theta_1)^{u-r+s}\left(-2\pi\right)^{r-s} 
  \mathcal{U}_{s, u} (\partial_\nu \Omega,\ldots) \,.
 \end{multline}
Finally, we consider
 \beq \label{defmathcalF}
  \left(2\pi i{\theta_{1}(\nu,\tau) \ov \theta'_{1}(0,\tau)} \right)^{2N}
  \mathcal{D}^r_s(\partial^2 \log\theta_1, \partial^3 \log\theta_1, \ldots)
  \mathcal{X}_\phi(\nu, \tau) \equiv \mathcal{F}^{r, s}(\nu, \tau)
 \eeq
and expand it in powers of
 \beq \label{vwvarib}
  v \equiv  e^{2i \pi \nu} \, \qquad w \equiv e^{2i \pi \tau} \,.
 \eeq
The coefficients of the power series in $v$ are
 \beq \label{coeffexpv}
  {1 \ov 2\pi i}\oint {dv \ov v} v^{m+N} \mathcal{F}^{r, s} = \chi^{r, s}_m (w).
 \eeq
The
expansion in powers of $v$ is valid inside an annulus 
between $v=1$ and $v=w$ (which are singular points, recall also
(\ref{nureg})). Remember, anyhow,
the discussion about the singularities in string one-loop amplitudes
after (\ref{nureg}): we found that the singularity at $v=1 \, (\nu=0)$
is actually not present for supersymmetric backgrounds (our
case)\footnote{It is instead understood in terms of the
  Fischler-Susskind mechanism when supersymmetry is absent.}. 
Therefore the integrand
of the one-loop amplitude has actually a zero at $v=1$.

We will show in section \ref{finitequ} that the coefficients
(\ref{coeffexpv}) obey a finite
difference equation,which will allow us to obtain their general form
and the dependence on the quantum numbers identifying the state
under consideration.
Before doing that, we complete the analysis of the one-loop amplitude
formula.

We further expand\footnote{Note that the coefficient of the series in
  powers of $v, w, \bar v, \bar w$ are real. This can be understood as
  follows. Let us 
  start with the bosonic part. The term $\left(2\pi
     i{\theta_{1}(\nu,\tau) \ov \theta'_{1}(0,\tau)} \right)^{2N}$ has
clearly an expansion with real coefficients. The polynomial
$\mathcal{P}$ depends on derivatives of the Green functions. The
state-vertex relation is 
$\alpha_{-n}^\mu \to i{\partial_\nu^n X^\mu \ov (n-1)!}$ and the
analog for right-moving quantities. 
By Wick's theorem, every term of $\mathcal{P}$ is given
by a product of two-point correlators of the form $\langle
i\partial_\nu^n i\partial_\nu^m G\rangle$ and analogues with
anti-holomorphic derivatives as well. The
  Green
  function on the torus for the scalar coordinates is $G
  =-{\alpha' \ov 2}\log|2 \pi i{\theta_1(\nu, \tau) \ov \theta'_1(0, \tau)}|^2
     +\alpha' \pi {(\nu-\bar\nu)^2 \ov 4\tau_2}$ 
which has an expansion of the form
     we are considering with real coefficients (the background charge
     part will be acted upon by the derivatives and become a constant term).
Every $\partial_\nu$ or $\partial_{\bar \nu}$ acting on $G$ brings down a factor
$i$. But every term in the polynomial $\mathcal{P}$ has an even number
of derivatives because of the level matching condition. Therefore the
coefficient of the expansion are real. The same is
valid for the antiholomorphic part. 

The fact that the fermionic contribution
has also an expansion in terms of real coefficients follows from the
same considerations, for example taking a bosonized form for the
world-sheet fermions and considering level matching as well as
conservation of fermion number.} as 
 \beq \label{coeffexptot}
 \chi^{r, s}_m(w) = \sum_p \chi^{r, s}_{m, p} \,w^p.
 \eeq
Note that $m+p \geq 0, \forall \,
m, p$.\footnote{This can be proven by observing that the torus
  amplitude can be divided into an invariant contribution under $\nu
  \to \nu+\tau$ , see section \ref{finitequ}, and a part which
  transform with periodicities related to the Neveu-Schwarz charges of
  the state. The terms we are expanding here are part of the invariant
  contribution. Furthermore, the torus 
integrand has a finite
limit for $\tau_2 \to \infty$ (that is $w \to 0$). Therefore for every
expansion term of the form $\chi_{m, p} w^p v^m$ under $\nu \to
\nu+\tau$ we have $w^p v^m \to w^{p+m} v^m$ and since the amplitude
must be finite for $w \to 0$, then $m+p \geq 0$.}
This will have a physical interpretation in (\ref{loopmasses}).

After expanding in the same way the anti-holomorphic part in power of
$\bar v, \bar w$,
we can finally write (\ref{oneloopfixedRN}) as
 \be \label{oneloopexpanded}
   T_{T^2}\!\! & \!\!= \!\!& \!\!
   g_s^2 \int {d^2\tau \ov \tau_2^{{d+1 \ov 2}+r+\tilde r-s}} 
   \int d^2\nu e^{-4(N+Q^2)\pi{\nu_2^2 \ov \tau_2}}
   \, \sum_{s, r, \tilde r} \sum_{p, m, \tilde p, \tilde m}
  \chi^{r, s}_{-m, p}\tilde\chi^{\tilde r, s}_{-\tilde m, \tilde p}
  \,\, \nonumber \\
  &&~~~~~~~~~~~~~~ \times e^{2\pi i p \tau +2\pi i m \nu}\,
   e^{-2\pi i \tilde p \bar\tau -2\pi i \tilde m \bar\nu}
  \nonumber \\
   & &~~~~~~~~~~~~~~\times \prod_{i=1}^{d_c-d}\frac{1}{R_{(i)}}\sum_{n^i, w^i}
  e^{\sum_{n_i,w_i}i\tau\left({n^i\ov R^i}+{w^iR^i\ov 4}\right)^2-i\bar\tau\left({n^i\ov R^i}-{w^iR^i\ov 4}\right)^2}
  \nonumber \\
  & &~~~~~~~~~~~~~~~~~~~~\times
   e^{-4\pi i \nu\sum_ip^i_L\left({n^i\ov R^i}+{w^iR^i\ov 4}\right)+4\pi i \bar\nu\sum_i p^i_R\left({n^i\ov R^i}-{w^iR^i\ov 4}\right)}. \nonumber \\  
 \ee
We see that the amplitude is now written as a sum of
different terms constituted by some coefficient factors 
$\chi^{r, s}_{-m, p}\tilde\chi^{\tilde r, s}_{-\tilde m, \tilde p}$ and
some integral factors
 \be \label{Hfact}
 H_{m, p, \tilde m, \tilde p,\{n^i, w^i\}} \!\! & \!\!= \!\!& \!\!
   \int_F {d^2\tau \ov \tau_2^{{d+1 \ov 2}+r+\tilde r-s}}
  e^{2i\pi\tau_1 \left(p-\tilde p+\sum_{i=1}^6 w_in_i\right)-2\pi\tau_2\left(p+\tilde p + \sum_{i=1} \left({2 n^2_i \ov R^2_i}+{w^2_i R^2_i \ov 8}\right)\right)}
 \nonumber \\
  && \int d^2\nu \,\,
  e^{-4(N+Q^2)\pi{\nu_2^2 \ov \tau_2}+2i\pi\nu_1 (m-\tilde
    m)-2\pi\nu_2(m+\tilde m-2N)} \nonumber \\
 &&~~~~~~~~~~~~ \times 
   e^{-4\pi i \nu\sum_ip^i_L\left({n^i\ov R^i}+{w^iR^i\ov 4}\right)+4\pi i \bar\nu\sum_i p^i_R\left({n^i\ov R^i}-{w^iR^i\ov 4}\right)}.
 \ee
Written in a compact way, it reads:
 \beq \label{toruscompactform}
  T_{T^2} = g_s^2\prod_{i=1}^{d_c-d}{1 \ov R_{(i)}}\sum_{n^i, w^i}
   \, \sum_{s, r, \tilde r} \sum_{p, m, \tilde p, \tilde m}
  \chi^{r, s}_{-m, p}\tilde\chi^{\tilde r, s}_{-\tilde m, \tilde p}\,\,
   H_{m, p, \tilde m, \tilde p,\{n^i, w^i\}}
 \eeq

The aim of the rest of this work is to determine the coefficient
($\chi^{r, s}_{-m, p}\tilde\chi^{\tilde r, s}_{-\tilde m, \tilde p} $)
and integral ($H_{m, p, \tilde m, \tilde p,\{n^i, w^i\}}$) factors and
finally compute the one-loop amplitude 
(\ref{toruscompactform}). 
We turn now to study these factors separately and in details.

\subsection{Integrals: real and imaginary part}
\label{realimagpart}

We start with the study of the integrals in (\ref{Hfact}).
Performing the integration over $\nu_1$, we find
 \beq
  m= \tilde m + \sum_i (n^i\mathcal{W}^i+w^i\mathcal{N}^i )
 \eeq 
where we have written
 \beq
 p^i_{L(R)}={\mathcal{N}_i \ov R_{(i)}}\pm {\mathcal{W}_iR_{(i)} \ov 4}.
 \eeq
However, recall that we limit ourselves to the case where the string
state does not have both non-zero
Kaluza-Klein and winding modes  at the same
time in a compact dimension. This implies $|p^i_L|=|p^i_R|$ and we
define $Q^i=p_L^i$.

In order to discuss the real and imaginary parts arising from the
integrals it is convenient to reduce them to a form which is pretty
similar to the one a field theory one-loop diagram
would have in Schwinger representation.
To do that, we change variables as
$\nu_2 \to \tau_2\nu_2$ and we find (we suppress indexes on $H$ to
neaten formulas) 
 \be \label{integralnutau}
  H & = & \int_F {d^2\tau \ov \tau_2^{{d+1 \ov 2}+r+\tilde r-s-1}}
  \int_0^1 d\nu_2 e^{2i\pi\tau_1 \left(p-\tilde p+\sum_{i=1}^6 w_in_i\right)}\\
  && \times e^{-4(N+Q^2)\pi\tau_2\,A(\nu_2)},
  \nonumber 
 \ee
where we have defined (here $i$ runs over the
compactified dimensions)
 \be \label{formAnu2}
 A(\nu_2) & = & 
 \text{{\small $\nu_2^2 + {1 \ov N+Q^2}\left(\!\!m\!-\!N\!-\!2\sum_i q_iQ^i\!\!\right)\nu_2$}} \nonumber \\
   &&\text{{\small $ +{1 \ov N+Q^2}({p+\tilde p -\sum_i n^iw^i\ov 2} + \sum_i q_i^2)$}} 
 \ee
with
 \beq
   q_i  = \left({n_i \ov R_{(i)}}+{w_i R_{(i)} \ov 4}\right).
 \eeq

Comparing this with a field theory one-loop diagram for a coupling
$\phi\varphi_1\varphi_2$, we find  
 \be \label{loopmasses}
  M_1^2 & = & 
  {p+\tilde p \ov 2}+\sum_i\left({n^2_i \ov R^2_{(i)}}+{w^2_i R^2_{(i)} \ov 16}\right)
  \nonumber \\
    \quad M_1^2-M_2^2 & = &
     -m+2\sum_i\left({n_i \ov R_{(i)}}+ {w_iR_{(i)} \ov 4}\right)Q^i
     -\sum_iQ^2_i, 
 \ee
where $M_1, M_2$ are the masses of the states (also virtual ones)
circulating in the loops.  

The function $A(\nu_2)$ is a parabola. For $m>N$ it is always
positive and therefore there is no imaginary part for the
integral. However when $m<N$ and for certain 
$p, \tilde p, {n_i, w_i}$, $A(\nu_2)$ is negative over a certain
range of $\nu_2$, which results in an imaginary part for the string self-energy.
In that case the integral as it stands is not well-defined and we have
to use analytical continuation for it. 

\begin{figure}[t]
 \begin{center}
 \includegraphics[height=6cm]{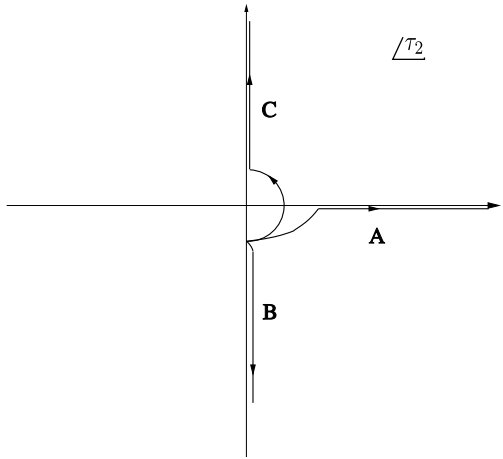}
\caption{{\small \sl{The analytical continuation for the integration
      over the modulus $\tau$.}}}  
 \label{pathcontinu}
\end{center}
\end{figure}

Differently from field theory,
we do not know the form of the amplitude for generic squared momenta
$p^2$, where $p^\mu$ is the momentum of the external lines in the diagram
in figure \ref{torustwopoint}. 
In fact, we can define it only for
specific {\em discrete} points $p^2=-(N+Q^2)$,
corresponding to the masses of on-shell string states, because only in
that case the vertex operators in the initial formula
(\ref{oneloopfixedRN}) are correctly defined. Furthermore,
note that off-shell string amplitudes are generally quite different form
their on-shell limit/counterpart (see \cite{Rastelli:2007gg} and
references therein). We therefore
cannot trust an analytical continuation to negative $(N+Q^2)$.

What we can do is instead to analytically continue the variable
$\tau$ (and therefore $\nu$). This is indeed the correct
procedure, since the integrand in $H$ is defined over continuous
regions of $\tau, \nu$. Also, see
\cite{Wilkinson:1989tb}, recall that physically the string world-sheet
propagator is defined as
 \beq
  \Delta = \int^{+\infty}_0 d\tau_2 \int_0^{2\pi} d\tau_1
  {1 \ov [\text{Invariances}]}
  e^{-i\tau_2(L_0+\tilde L_0-i\epsilon)-i\tau_1(L_0-\tilde L_0)}
 \eeq
and we usually consider its Euclidean continuation $\tau_2 \to
-i\tau_2$ (from $A$ to $B$, in figure \ref{pathcontinu}). However,
this continuation is possible only if the eigenvalues of $(L_0+\tilde
L_0-i\epsilon)$ are positive defined. This is not the case when the
system is unstable and we can have decay, and this leads to the
aforementioned imaginary parts.

The correct analytical continuation is, in that case,  $\tau_2 \to
i\tau_2$ (from $A$ to $C$ in figure \ref{pathcontinu}). 
In performing it\footnote{Note that the total
  amplitude, when analytically continued remains convergent since the
  eta and theta functions are still convergent (only a
  finite number of terms in their expansion in powers of $w=e^{2\pi i
    \tau}$ will be affected by the continuation, namely those powers
  with exponent $0<n<N$ only).},  we obtain an
imaginary part and a real part. Indeed the new path $C$ of integration for
$\tau_2$ can
be divided into a contribution from the imaginary axis and a
semicircle shrinking
around $\tau=0$. The imaginary part comes from the integration around
the semicircle, as was discussed and obtained in
\cite{{Mitchell:1988qe}, Wilkinson:1989tb}.

The real part comes from the
integration of $\tau_2$ along the axis. In terms of $\tau$,
the integration is over the complex plane, dividing out the
invariances of the metric, namely modular invariance. As we already
mentioned this leads to the 
definition of fundamental regions of integration. 

Note that in the
usual operatorial formulation of torus amplitudes in string theory
\cite{Green:1987mn},
the restriction of the integration
region over a fundamental domain is sometimes inferred by checking invariance of
the one-loop amplitude once it is already analytically continued to
$\tau_2 \to -i\tau_2$, but the amplitude is formally divergent (due to
the imaginary part)! 

The true physical reason for the restriction of the integration
region is instead the necessity of dividing out the
gauge invariances of the torus metric
(see Polchinski volume 1, chapter 5)
 \beq
  ds^2= |d\sigma_1^2+\tau d\sigma_2^2|.
 \eeq
and this must be taken into account by any
analytical continuation.

We will now present first the formulas for the imaginary part, and
then those for the real one.

\subsubsection{The imaginary part}
~~
\nopagebreak

For specific values of $m(<N), \,p,\, \tilde p$ and $\{q^i\}$, we can
write (\ref{formAnu2}), as
 \beq \label{formAnu2realroots}
 A(\nu_2) =  (\nu_2-y_-)(\nu_2-y_+) 
 \eeq
with two real root $y_+, \, y_-$ given by ($i$ runs over the
compactified dimensions)
 \be \label{realroots}
 y_\pm & = & \text{{\small $
-{1 \ov 2(N+Q^2)}\left(m-N-2\sum_i q_iQ^i\right)$}} \\
 &&\text{{\small $
\pm \sqrt{\left({m-N-2\sum_i q_iQ^i\ov 2(N+Q^2)}\right)^2-{1 \ov
    (N+Q^2)}\left( {p+\tilde p -\sum_i n^iw^i \ov 2}+\sum_iq_i^2\right)}$}}. \nonumber 
 \ee
Here $n, p, {n_i, w_i}$ run only
over values that make the square root real.

From (\ref{integralnutau}), after the analytical continuation and
integration over $\tau_1$,
 \beq
  \text{Im}(H)=
  \int^{y_+}_{y_-} d\nu_2 
 {\sin(\pi(p-\tilde p+\sum_i n_iw_i)\ov\pi(p-\tilde p+\sum_i n_iw_i)}
 \pi 
  {(4(N+Q^2)\pi |A(\nu_2)|)^{({d+1 \ov 2}+r+\tilde r-s-2)} \ov \Gamma(({d+1 \ov 2}+r+\tilde r-s-1))}\, , 
 \eeq 
where $\nu_2$ is integrated over the range where $A(\nu_2)$ is
negative (see (\ref{formAnu2realroots})).

As we can see, the imaginary part is non-zero only for 
 \beq \label{levelmatch}
  p-\tilde p+\sum_{i \epsilon \text{comp}} n_iw_i = 0
 \eeq 
which ensures that the emitted string states satisfy the level
matching condition.

Integrating over $\nu_2$ and inserting back the result in
(\ref{oneloopexpanded}), we obtain
 \beq \label{imaginarypartform}
 \text{Im}(T_{T^2})= g_s^2\prod_{i=1}^{d_c-d}\!\!{1 \ov R_{(i)}}\!\!
  \sum_{\{n^i, w^i\}}
  \sum_{s, r, \tilde r} \sum_{m, p} 
  \chi^{r, s}_{-m, p}\,\tilde\chi^{\tilde r, s}_{-\tilde m, \tilde p} \,
   \,{\pi \ov 2\sqrt{N+Q^2}}
   {(4\pi \varpi^2)^{{d+1 \ov 2}+r+\tilde r -s-{3 \ov 2}} \ov
     \Gamma({d \ov 2}+r+\tilde r -s)}
 \eeq
where we have defined ($i$ runs over the compactified dimensions)
 \be \label{emitmom}
  \varpi & \equiv & \sqrt{N+Q^2}\,{y_+ - y_- \ov 2} \\
    & = &\text{{\small$
   \sqrt{\left({m-N-2\sum_i q_iQ^i \ov 2\sqrt{N+Q^2}}\right)^2-\left({p+\tilde p -\sum_i n^iw^i \ov 2}+ \sum_i q_i^2\right)}$}} \nonumber
 \ee
and $\tilde p$ is given by (\ref{levelmatch}). 

The result (\ref{imaginarypartform}) is analogous to those, for example, in
\cite{Chialva:2003hg}.   

\subsubsection{The real part} \label{realpartgen}
~~
\nopagebreak

We must consider, defining it by analytical continuation where necessary,
 \be \label{integralnutaufin}
 H & = & \int_F {d^2\tau \ov \tau_2^{{d+1 \ov 2}+r+\tilde r-s}}
  e^{2i\pi\tau_1 \left(p-\tilde p\right)-2\pi\tau_2\left(p+\tilde p-\sum_i n^iw^i + 2 \sum_iq_i^2\right)}
 \nonumber \\
  && \int d^2\nu 
  e^{-4(N+Q^2)\pi{\nu_2^2 \ov \tau_2}-4\pi\nu_2(m-N-2\,\sum_i q_iQ^i)}\,  .
 \ee
The integral does not exist in finite form. In order to estimate
correctly its value, we need to know better
the asymptotes of the integrand in the various regions of
integration. These will depend on the values of $m, p, \tilde p,
\ldots$ and of course on $N, Q$. 
It is therefore convenient to discuss them once we will have
a more precise knowledge of the other
ingredients of the one-loop amplitude (the expansion coefficients 
(\ref{coeffexptot})) that will enable us to individuate the dominating
contributions in terms of $m, p, \tilde p,
\ldots$.

In any case, by inspecting
(\ref{integralnutaufin}) we can start arguing how the integral
factor would affect, and suppress, certain contributions to the total
sum over $p, \tilde p, m, n^i, w^i$. The dominant contributions should
give the correct asymptotic behavior of the whole amplitude.
 
We observe that terms with large $p, \tilde p, q^2$ would be generally
suppressed by the exponential in the first line of
(\ref{integralnutaufin}) and force $\tau_2$ to the lower limit of its
integration region, which in turn would suppress the contribution due
to the exponential in the second line (in particular due to the first
term in the exponential). Analogously, contributions with $m \gg N$
will be similarly suppressed.

Note that, due to the identification  
(\ref{loopmasses}), the channel(s) where $p, \tilde p = 0$ 
correspond to the case where at least one of the two states
running in the loop is massless or a pure
Kaluza-Klein/winding mode\footnote{The condition we give is 
  for $M_1^2=0$. The condition for $M_2^2=0$ is $m=-{p+\tilde p \ov2}$, 
  but since the amplitude is invariant under $M_1^2 
  \leftrightarrow M_2^2$ (one can check it comes from the invariance
 under $\nu \leftrightarrow \tau-\nu$
  of the periodical -quantum- contribution to the torus amplitude, see
  section \ref{finitequ}), we need only to treat the case
 $p =\tilde p= 0$, which will turn out to be simpler to deal with,
  given the formulas 
  (\ref{chir}) for the coefficients. \label{masslesscond}}. 
That massless interactions (namely the gravitational one)
dominate the mass shift for very massive string is expected on general
reasons\footnote{The rationale behind
this idea is that
the gravitational interaction grows with the mass and therefore
becomes the dominant one for very massive states.} (\cite{HorPolSelf,
  DamVenSelf})  
and will be further discussed in section
\ref{dominantchanneldiscuss}. 

Evidently, we need to check if
the more or less  
suppression due to the integral factors in the amplitude is
compensated or not by the behavior of the coefficient factors.
Therefore our 
argument so far is not sufficient to determine the favored
channels of interaction.

We will now indeed turn to the study of the
coefficients in the expansion of the one-loop amplitude.

\subsection{Coefficients}\label{coeffgen}

In this section we are going to obtain the coefficients of the
expansion of the one-loop amplitude (\ref{toruscompactform}), as
defined in section 
\ref{subsec_exponeloop}, formulas (\ref{coeffexpv}, \ref{coeffexptot}).  

\subsubsection{The finite difference equation} \label{finitequ}

We will exploit the transformation properties of the one-loop
amplitudes under some symmetry transformations.
This will enable us to write a system of finite difference equations
that the coefficients 
(\ref{coeffexpv}) must obey. By solving it
we will obtain the form of the coefficients. The discussion goes in
parallel for both $\chi^{r, s}_m$ and 
$\tilde \chi^{\tilde r, s}_{\tilde m}$,
therefore we will treat in details the first ones.
 
\begin{figure}[t]
 \begin{center}
\includegraphics[height=6.6cm]{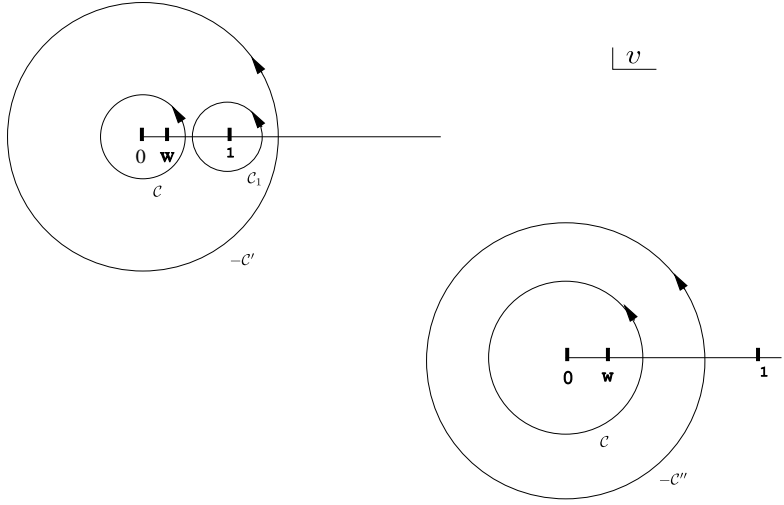}
\caption{{\small \sl{The contours and relevant deformations
      involved in the derivation of
      the finite difference equation for the expansion coefficients of
      the one-loop amplitude.}}}  
 \label{contours}
\end{center}
\end{figure}

Beside the contour $\mathcal{C}$
we used in the definition (\ref{coeffexpv}), we consider
two different contours\footnote{The idea of exploiting specific transformations
  laws 
  in order to obtain general equations constraining string amplitudes,
  when expanded in suitable power series, was first used in
  \cite{Manes:2001cs} on partial inclusive tree-level amplitudes. 
  Here we apply it to
  one-loop amplitudes, modifying it where useful. Ours 
  analysis of the equations and their
  solution is different from that in \cite{Manes:2001cs}.} 
$\mathcal{C}^\prime, \, \mathcal{C}^{\prime\prime}$ (see figure
\ref{contours}),    
related by the transformations $v'=vw^{-1}, \, v^{\prime\prime}=wv^{-1}$.

Let us understand how the integrand of the one-loop amplitude
(\ref{oneloopfixedRN}) behaves
under the transformations that relate the various loops integrations.
Knowledge of some properties of the integrand in
(\ref{oneloopfixedRN}) will be useful at this point.
In particular, under the torus periodicities
 \beq \label{nuperiod} 
  \nu \to \nu \pm 1, \qquad \nu \to \nu \pm \tau,
 \eeq
the compactified coordinates transform as
 \be
  X^i(\nu \pm 1) & = &  X^i(\nu) + 2\pi R_{(i)}w^i \\
  X^i(\nu \pm \tau) & = &  X^i(\nu) + 2\pi R_{(i)} l^i
 \ee
where $l^i$ is the dual (in the sense of Poisson resummation) of the
Kaluza-Klein number $n^i$. The path integral can be computed as usual
by writing $X^i$ as a classical solution with the correct periodicity
plus a quantum piece with periodic boundary conditions.

Since our states are not excited in the compactified
dimensions, the quantum (periodic) part is the only one that enters in
$\mathcal{F}^{r, s}(\nu, \tau)$, 
defined in (\ref{defmathcalF}). Invariance under
(\ref{nuperiod}) of the total periodic
contribution to the torus amplitude implies that it
transforms as
 \beq
 \mathcal{F}^{r, s}(\nu\pm\tau, \tau) = 
   e^{\mp i 4 N \pi \nu} e^{-i 2 N \pi \tau} \mathcal{F}^{r, s}(\nu, \tau). 
 \eeq

By opportunely deforming the various contours, picking up the residues
at the singular points $v=0, 1$, we find
 \be \label{eqCprime}
  {1 \ov 2\pi i}\oint_{\mathcal{C}^\prime}{dv^\prime \ov v^\prime}v^{\prime\, m+N}
 \mathcal{F}^{r, s}(v', w) & = &
  {1 \ov 2\pi i}\oint_{\mathcal{C}^\prime}{dv \ov v}{v^{m+3N} \ov w^{m+2N}}
 \mathcal{F}^{r, s}(v, w)   \nonumber \\
   & = &
  {1 \ov 2\pi i}\oint_{\mathcal{C}}{dv \ov v}v^{m+N} \mathcal{F}^{r, s}(v, w) +
 \mathcal{R}^{r, s}_m
 \ee
with  
 \beq \label{residuedef}
  \mathcal{R}^{r, s}_m = 
  {1 \ov 2\pi i}\oint_{\mathcal{C}_1}{dv \ov v}v^{m+N} \mathcal{F}^{r, s}(v, w) \, ,
 \eeq 
where $\mathcal{C}_1$ is a circuit around $v=1$.

Similarly\footnote{Here we also used the invariance of
  $\mathcal{F}^{r, s}(\nu, \tau)$ under
under 
  $ \nu \to -\nu$,
due to the fact that the whole amplitude is invariant under the modular
transformations 
 $ \nu \to {\nu \ov c \tau +d}, \quad \tau \to {a \tau +b \ov c \tau +d},
  \quad ad-cb =1$
and we can choose $a=d=-1, b=c=0$.}
 \be \label{eqCthird}
  {1 \ov 2\pi i}\oint_{\mathcal{C}^{\prime\prime}}{dv^{\prime\prime} \ov v^{\prime\prime}}
     v^{\prime\prime\, m+N} \mathcal{F}^{r, s}(v^{\prime\prime}, w) & = &
  {1 \ov 2\pi i}\oint_{\mathcal{C}^{\prime\prime}}{dv \ov v}{v^{N-m} \ov w^{-m}} \mathcal{F}^{r, s}(v, w) \nonumber \\
    & = &
  {1 \ov 2\pi i}\oint_{\mathcal{C}}{dv \ov v}v^{m+N} \mathcal{F}^{r, s}(v, w) .
 \ee
We compute the residue $\mathcal{R}^{r, s}_m$ in the appendix
\ref{residuecomp}, with 
the result
 \beq \label{residueRresults}
  \mathcal{R}^{r, s}_m = 0 \,.
 \eeq
Using the definition (\ref{coeffexpv}), equations 
(\ref{eqCprime}, \ref{eqCthird}) read
 \beq \label{findiffeq}
  \begin{cases}
   \chi^{r, s}_{m+2N}(w) = w^{m+2N}\,\chi^{r,s}_{m}(w) 
   \\
   \chi^{r, s}_{-m}(w) = w^{-m}\chi^{r, s}_{m}(w)\, .
  \end{cases}
 \eeq

\subsubsection{The solution of the finite difference equation}

We will now solve the system (\ref{findiffeq}). First, we note that
the integrand of the torus amplitude has a finite result for $w
\to 0$ ($\tau_2 \to \infty$). This implies that the expansion in
series of $w$ for all
coefficients $\chi^{r, s}_m(w)$, for both positive and negative
$m$, can have only positive or null powers.  
Then, consider the last equation of (\ref{findiffeq}): for
$m>0 $, being $w^{-m}$ a negative power of $w$, it must be 
 \beq \label{positivewcoeff}
 \chi^{r, s}_{m}(w) \sim a_m w^{m}.
 \eeq
The general solution of (\ref{findiffeq}) is
 \begin{align}
 \chi^{r, s}_{\pm m}(w) =  w^{A N(A \pm 1)+\eta(A+{1 \pm 1 \ov 2})}\chi^{r, s}_{-\eta}(w)
  \label{chir}
 \end{align}
with
 \beq
 0 \leq \eta \leq  2N, \qquad m=2AN+\eta.
 \eeq
The solutions depend
 on some unknown coefficients $\chi^{r, s}_{-\eta}, \, 0 \leq \eta \leq
 2N$. They contain the information on the specific state (pure
 or mixed) whose one-loop two-point function we analyze. Not all of
 them are independent, in fact using (\ref{findiffeq}), we see that only the
 coefficients with $0 \leq \eta\leq N$ are such. 

The system
 (\ref{findiffeq}) by itself is not enough to determine them, but
 these coefficients (with one exception) are those which participate
 in the imaginary 
 part of the amplitude.  The only exception is the terms $\eta=N$, for
 which the amplitude does not have an imaginary part. This is the only
 coefficient we cannot fully determine. We will deal with the
 contribution $\eta=M$ in more details in section \ref{avmassshift}.

We can then compute the coefficients with $0 \leq \eta \leq N-1$ from the
 tree-level amplitudes for the decays of our initial state into two
 string states. Unfortunately this again would in principle require
 the knowledge 
of the form of all string vertex operators, which we do not possess.

We turn now to the
discussion of the specific case we are interested in:
the average closed string mass shift. We will see that it is possible
to estimate what are the dominant contributions to the one-loop
self-energy and compute exactly the relevant coefficients and integral
factors in that case.

\section{The average mass shift for closed string
  states}\label{averagemassshift} 

The aim of this section is to obtain the average mass shift for closed
string states at three-level squared mass $M^2_0=N+Q^2$ for fixed $N$
and $Q$. To the reader's 
convenience, 
let us summarize
here our achievements so far and let us anticipate what we are going
to do now. 

In the previous sections,
we have studied in details the one-loop string amplitude, individuating
its imaginary and real parts. The amplitude has been written in
terms of a sum over coefficients and integral terms. Formulas for the
integrals have been given, both for the imaginary and the real part. 
Formulas for the coefficients have been given as well, in
terms of a subset of such coefficients that strictly depend on the
state under consideration.
On the other hand, we have proven in section \ref{realimagpart}
that this remaining subset of coefficients (a part from one) can
actually be determined by comparison to three-level amplitudes
(decay rates). 

We are going now to 
 \begin{itemize}
  \item define the relevant amplitudes and decay rates in order to
    obtain the coefficients which are still unknown in (\ref{chir}). 
  \item obtain the asymptotic behavior of the whole on-loop amplitude
    for large initial $N=M_0^2-Q^2$ by studying the different kinds
    of contributions and individuating the asymptotically dominant ones.
 \end{itemize}

\subsection{Definition of the relevant decay rates}

We define here the tree-level computations useful to obtain the
remaining unknown expansion coefficients in (\ref{chir})
for the one-loop average string
amplitude.
They have the form:
 \beq \label{gendecayaverage}
 \Gamma = g_s^2 \mathcal{P}\sigma^{\text{av}}_L \times \sigma^{\text{av}}_R
 \eeq 
 where
 \begin{itemize}
 \item $\mathcal{P}$ is the {\em phase space}, including the sum over
 the Neveu-Schwarz charges of the final states and the normalization
 factor for the amplitude
 \item 
   \beq \label{sigmaRdef}
   \sigma_{L}^{\text{av}} = {1 \ov \sqrt{G_c(N)}}\sum_{\phi_{|_N}}
   \sum_{\phi', \,\zeta}
     |\langle\phi'_L| V_{\zeta,\,L} |\phi_L\rangle|^2
   \eeq 
   is the sum over the final states\footnote{Our notation for closed
     string states here is 
 $|\phi\rangle = |\phi_L\rangle |\phi_R\rangle$.} $\phi', \zeta$ 
of the modulus square of the decay amplitudes for the state(s) $\phi$,
averaged over physical states
$\phi$ at mass level $N=N_R=N_L$ carrying fixed charges 
$Q^i=Q^i_L$,\, $|Q^i_R|=|Q^i_L|$. \,\,
$\sigma_R$ is obtained
 substituting to 
   the left-moving quantities the right-moving ones.
 $G_c(N)$ is the degeneracy of closed string states at
   mass-level $N$.
 \end{itemize}

In order to compare the sum over tree-level decays with the imaginary
part of the one-loop amplitude, it is convenient to reorganize the sum
over states $\phi',\, \zeta$ as a double sum over states at a given mass-level
$N',\, \aleph$ and all possible mass-levels:
 \beq
  \sum_\zeta = \sum_\aleph\sum_{\zeta_{|_\aleph}} \qquad 
  \sum_{\phi'} = \sum_{N'}\sum_{\phi'_{|_{N'}}} \,\, ,
 \eeq
and define
 \beq 
  \sigma^{\aleph, N'}_{L} \equiv \sum_{\zeta_{|_\aleph}, \phi_{|_N}} 
   {1 \ov \sqrt{G_c(N)}} {1 \ov 2\pi i}\oint {dz' \ov z^{\prime\,N'+1}}
  \langle\phi_L|V^\dag_{\zeta,\,L}(1)\,z^{\prime\,\hat N_L}\,V_{\zeta,\,L}(1)|\phi_L\rangle.
 \eeq
having used a mass-projector (\cite{AmRus})
 \beq
   \rho_{N'}={1 \ov 2\pi i}\oint {dz' \ov z^{\prime\,N'+1}}z^{\prime\,\hat N_L},
 \eeq
such that
 \beq
  \sum_{\phi'_{|_{N'}}} |\phi'_L\rangle\langle\phi'_L| =
   \sum_{\phi'_{physical}} \rho_{N'}|\phi'_L\rangle\langle\phi'_L| \,.
 \eeq
The average case can be discussed in terms of  a density matrix
 \beq \label{averagecasedensmat}
 \rho_{\text{av}} = 
   \sum_{\phi_{|_\text{physical}}} c_{\phi} |\phi\rangle\langle\phi|.
 \eeq
defined by
 \beq \label{averagecasedensmatcoeff}
  \begin{cases}
    c_\phi \equiv G_c(N) & \text{for $\phi$ at
    mass-level $N$} \\
    c_\phi \equiv 0 & \text{for $\phi$ not at
    mass-level $N$} \qquad.
  \end{cases}
 \eeq
In our case, as we said, it is also $N_R=N_L=N$ (see
(\ref{massshellcond})).

We can see then that the average is just a particular case of a more
general scenario and that our approach 
can be extended to other cases and 
used for obtaining other statistical informations using different
density matrices representing different mixed states.

Let us concentrate now on the average case only.
Its density matrix can be re-written as
$\rho_{\text{av}}=\rho_N\tilde \rho_N$ with
 \beq
   \rho_N={1 \ov \sqrt{G_c(N)}}{1 \ov 2\pi i}\oint {dx \ov x^{N+1}}x^{\hat N_L}, 
 \eeq
and we obtain
 \be \label{sigmatwoindex}
  \sigma_{L}^{\aleph, N'}  & = & \sum_{\zeta_{|_\aleph}, \phi_{|_N}} 
   {1 \ov \sqrt{G_c(N)}} {1 \ov 2\pi i}\oint {dz' \ov z^{\prime\,N'+1}}
  \langle\phi_L|V^\dag_{\zeta,\,L}(1)\,z^{\prime\,\hat N_L}\,V_{\zeta,\,L}(1)|\phi_L\rangle 
    \nonumber \\
   & = & {1 \ov \sqrt{G_c(N)}}{1 \ov (2\pi i)^2}
    \sum_{\zeta_{|_\aleph}} \oint {dx \ov x^{N+1}}\oint {dz' \ov z^{\prime\,N'+1}}
      \text{tr}[V^\dag_{\zeta,\,L}(1)V_{\zeta,\,L}(z')(xz')^{\hat N_L}] 
   \nonumber \\
  & = & {1 \ov \sqrt{G_c(N)}}{1 \ov (2\pi i)^2}
    \sum_{\zeta_{|_\aleph}} \oint {dz \ov z^{N+1}}\oint {dz' \ov z'}z^{\prime\,N-N'}
      \text{tr}[V^\dag_{\zeta,\,L}(1)V_{\zeta,\,L}(z')\,z^{\hat N_L}] \nonumber
      \\ 
 \ee
where in the last line we have changed variables to
 \beq
  z \equiv xz'\, .
 \eeq
Here $\hat N_L$ is the left-moving mass-level operator\footnote{We will always
  distinguish an operator from its value by mean of a $\hat{}$\,.}. We
can write $\sigma_R$ in the same way using $\hat N_R$.

The last line of formula (\ref{sigmatwoindex}) is our main formula:
$\sigma^{\aleph, N'}_{L(R)}$ written in this way can be computed in operator
formalism in the same way as
the one-loop two point function for on-shell vertex operators $V^\dagger_\zeta,
V_\zeta$, just without integrating over the zero modes, without
spin-structure one-loop signs and
projecting at the end on some mass-level. This
ensures that only physical states enter in the trace and therefore
that these on-shell amplitudes are well-defined in string theory.

Let us fix clearly the notation for future use (repeated index is summed):
 \begin{itemize}
  \item the initial state has momentum
 $p^\mu$, mass-level $N$, Neveu-Schwarz 
 charges $Q^i=Q^i_L$,\, $|Q^i_R|=|Q^i_L|$, from (\ref{massshellcond})
 and tree-level squared mass
     $M^2_0 = N+Q^2, \,\,Q^2=Q^iQ_i$, 
 \item the emitted state $\zeta$ has momentum $k$,
   mass-level $\aleph$, charges $q^i=q_L^i$ (and $q_R^i$) 
    and tree-level squared mass 
   $M^2_2 = \aleph+q^2, \,\,q^2=q^iq_i$, 
 \item the other final state has momentum $p'$, mass-level
 $N'$, charges $Q^{\prime \,i}=Q^{\prime \,i}_L$ (and $Q^{\prime \,i}_R$) 
  and tree-level squared mass 
   $M_1^{2} = N' + Q^{\prime\, 2}, \,\,Q^{\prime\, 2}=Q^{\prime \,i}Q'_{i}$.  
 \end{itemize}

\vskip 0.2cm
\noindent
The phase space then reads
 \beq \label{phasespacgeneral}
  \mathcal{P}  = c
     {2\pi^{{d \ov 2}} \ov \Gamma({d \ov 2})}\prod_{i=1}^{dc-d}{1 \ov R_{(i)}}
     \sum_{\{q_i\}} {1 \ov N+Q^2} 
   \sqrt{\left({N-N'+\aleph+2q^iQ_i \ov 2\sqrt{N+Q^2}}\right)^2\!\!-\aleph\!\!-\!\!\sum_i q^iq_i}^{D-3} \, ,
 \eeq
where $c$ is the overall normalization, and so we can write
 \be \label{decayfinform}
  \Gamma\!\!\!\! & = &\!\!\!\! cg_s^2
     {2\pi^{{d \ov 2}} \ov \Gamma({d \ov 2})}\prod_{i=1}^{d_c-d}{1 \ov R_{(i)}}
     \sum_{\{q_i\}, \aleph, N'} 
   \sqrt{\left({N-N'+\aleph+2q^iQ_i \ov 2\sqrt{N+Q^2}}\right)^2\!\!-\aleph\!\!-\!\!\sum_i q^iq_i}^{D-3} \nonumber \\
   & & \times {1 \ov N+Q^2}\,\sigma_L^{\aleph, N'}\,\sigma_R^{\aleph, N'}
 \ee

We can compare this with the one-loop result for the imaginary part,
given in (\ref{imaginarypartform}, \ref{emitmom}).

By using
 \beq \label{imanddec}
  {4 \pi \ov M_0}\text{Im}(T_{T^2}) = \Gamma.
 \eeq
and we obtain
 \beq  \label{coefffromtree} 
  \sum_{r, \tilde r, s} {4^{r+\tilde r-s} \ov \Gamma({d \ov 2}+r+\tilde r-s)} 
    \chi^{r, s}_{-m, p} \tilde \chi^{r, s}_{-m, p} = c
   {2^{3-D} \ov \pi \,\Gamma({d \ov 2})}
   \sigma_R^{p, m+p}\, \sigma_L^{p, m+p}\, ,
 \eeq
where we have used (\ref{levelmatch}) in the case $q^i_R=q^i_L$ and
(\ref{loopmasses}).

\subsection{Dominant channels of interaction}\label{dominantchanneldiscuss}
 
We will try now to argue which, among the possible contributions to
the one-loop amplitude, for different $p, \tilde p, m, \ldots$, are
the dominant ones. These should provide us with the asymptotically
correct result for the self-energy.

By looking at (\ref{averagecasedensmat}, \ref{averagecasedensmatcoeff}), 
an apparent advantage seems to come from the fact that the integrand
of the one-loop amplitude for the average case
(\ref{oneloopfixedRN}) will be weighted by a
factor\footnote{It will indeed enter the terms $\mathcal{P}_{\rho_{\text{av}}},
\mathcal{X}_{\rho_{\text{av}}}, \bar{\mathcal{X}}_{\rho_{\text{av}}}$
in the amplitude (see section \ref{oneloopform}), which
depend on the density matrix coefficients $c_\phi$.} 
$G_c(N)^{-1}$, where $G_c(N)$ is the string degeneracy at level
$N$. This would
strongly suppress the coefficients of many contributions
(see (\ref{coefffromtree})) 
by a factor $\sim e^{-2\pi\sqrt{N\,(d-1)}}$ for large
$N$. Non-suppressed contributions are those for which this factor is
compensated. However, simple arguments
can show that this can in fact happen,
because of the summation
over all decay channels. These arguments are presented in the appendix
\ref{eurargkin} for what concerns the contribution with non-zero
imaginary part. 
 
At the same time, these same remarks 
exclude that coefficient factors are exponentially enhanced.

On the other hand, as we already said, different
considerations point at individuating the dominating contribution to
the mass shift in those channels of interactions where at least one of the two
states running in the loop is massless or possibly a pure
Kaluza-Klein/winding mode. 

We call these {\em massless
  interactions}, while the term {\em massive} will represent those for
which both  
  states running in the loop have non-zero mass level, that is they
  are massive form a ten dimensional point of view.   

On general ground, gravitational
interactions 
(where one of the states running in the loop is a graviton, dilaton,
Kalb-Ramond field or in general a superposition of them) are supposed to
dominate the self-energy of (very) massive strings since the
interaction grows with the mass (\cite{HorPolSelf, DamVenSelf}).

Another argument for this comes from the direct evaluation
and estimate of the 
asymptotic dependence for large initial mass $M$ of the contributions
coming from massless and
massive channels.
Our expression for the one-loop amplitude (\ref{toruscompactform})
allows to easily distinguish the two. We will now discuss the latter
and then the supposedly dominant massless channels.

It is difficult to determine exactly the contribution of the total of the
massive channels, but we can estimate it, when the massive states 
running in the loop have $1 \ll M_{1, 2} < M$. Indeed, from
(\ref{coefffromtree}) the relevant coefficient factors in those cases
can be obtained from the semi-inclusive decay rate  $M \to
M_1+M_2$. We can estimate 
the latter in the following way. 

A closed string can
decay in two massive strings
only when two points on the string get in contact. The
first point can be placed everywhere on the string, and therefore, for a
highly massive string behaving like a random walk \cite{chialvasize},
the probability of decay is directly proportional to the (average)
length\footnote{We use the notation introduced in the previous section.}
$N^{{1 \ov 4}}$ of the string. For $M_{1, 2} \gg 1$ also the
final decay products can be considered as random walks of length
$\aleph^{{1 \ov 4}}, N^{\prime\,{1 \ov 4}}$. 
Then, the probability for two points at a distance
$\text{min}(\aleph^{{1 \ov 4}}, N^{\prime\,{1 \ov 4}})$ to meet is directly
proportional to the relative volumes 
$\left({N \ov \aleph\, N'}\right)^{{D-1 \ov 4}}$ in $d$ spatial
dimensions.

Therefore (we use a
sloppy notation here)
 \beq
  \sum_{r, \tilde r, s} \chi^{r, s} \tilde \chi^{\tilde r, s} \sim 
  M\, \Gamma_{M, M_1, M_2} \sim 
   N^{{1 \ov 4}}\, \left({N \ov \aleph\, N'}\right)^{{D-1 \ov 4}} \,.
 \eeq
If we write $N' = x\,N, \,\aleph \sim (1-x)\,N$, then
 \beq
  \sum \chi \tilde \chi \approx N^{{2-D \ov 4}} (x-x^2)^{{1-D \ov 4}}\,.
 \eeq
We see that the coefficient factors are
maximized when either $N'$ or $\aleph$ are small, that is when
nearly in the massless case. 

At the same time, as  we have already
discussed at the end of section \ref{realimagpart}, generically all
massive channels are  suppressed by the integral factor(s) in the
amplitude (see
(\ref{integralnutaufin}) for $m, p, \tilde p \neq 0$). In particular
off-shell massive states will be strongly exponentially suppressed by
the integrals.
 
Therefore the only possibly non-negligible contributions, compared to
those of the massless channels that we will discuss in a moment (see
(\ref{estimsigmag})) appear to occur 
when one of the states running in the loop is nearly
massless. We will consider them a (possibly small)
correction to the result given by computing the asymptotic behavior of the
massless channels, which will give us the correct asymptotes for the
whole amplitude.

We will give further evidence for the suppression of channels other
than the massless ones
in section \ref{avmassshift}. We will also show that
indeed  the dominating 
contribution comes from the 
gravitational and not
other forms of long distance interaction (provided by the other string
massless states). We will also see that generally
Kaluza-Klein and winding modes are suppressed, and only the lowest ones
can be sufficiently excited.

We are aware that the conclusion that the total contribution of
massive channels of interactions is really asymptotically subdominant
with respect to the massless ones' does not come from a complete and exhaustive
calculation, but we believe we have provided
some more arguments supporting this, beyond the physical intuition
regarding the dominance of gravitational interactions for very large masses.
However, even adopting a more conservative and minimal point of
view, we will be able at least to obtain the correct corrections due
to the massless channels, therefore verifying previous approximated
result in the literature, which assumed the predominance of
massless (gravitational) interactions from the outset.

Looking back at (\ref{loopmasses}), we see that we have massless or pure
Kaluza-Klein/winding
modes circulating in the loop when\footnote{See also note \ref{masslesscond}.}
 \beq
  p =\tilde p= 0,
 \eeq
which means that their contribution is given by the term $w^0$
in the power series of $\chi^{r, s}_m(w)$. 

Therefore, taking the limit
$w \to 0$, we find\footnote{See also appendix \ref{residuecomp} for a
  discussion of this result.}
 \beq \label{masslesslimitcoeff}
 \begin{cases}
 \chi^{r, s}_m(w) \to 0 & \text{for $m > 0 $ and $m < -2N$} \\
 \chi^{r, s}_{-m}(w) \to \text{unknown} & \text{for $0\leq m \leq N $} \\
 \chi^{r, s}_{-m}(w) = \chi^{r, s}_{m-2N}(w)  
   & \text{for $N < m \leq 2N $}\, ,
 \end{cases}
 \eeq
using (\ref{findiffeq}, \ref{positivewcoeff}, \ref{chir}). The same
occurs for $\tilde \chi^{\tilde r, s}_{\tilde m}$.

The unknown terms are those we will focus on in
the following. Since they contribute to the imaginary part (a
part from $\chi^{r, s}_{-N},\, \tilde \chi^{\tilde r, s}_{-N}$ which
we will consider later on), as 
we said we can obtain them form a tree-level computation by using
formula (\ref{coefffromtree}).

In particular, we will compute
 \beq \label{gammadom}
  \Gamma  =
   g_s^2  \sum_{N'}\mathcal{P}\,\sigma^{0, N'}_L \, \sigma^{0, N'}_R\, .
 \eeq
Performing 
the sum over all states $\zeta$ at mass level $\aleph=0$ means to
compute the quantity
 \be \label{sigmasdom}
  \sigma^{0, N'}_L \times \sigma^{0, N'}_R & = &
  (\sigma^{g, N'}_L\sigma^{g, N'}_R+\sigma^{s, N'}_L\sigma^{s, N'}_R + \sigma^{g, N'}_L \sigma^{s, N'}_R+ \sigma^{g, N'}_R \sigma^s_L)
   \nonumber \\
  & = &
    (\sigma^{g, N'}_L+\sigma^{s, N'}_L) \times 
   (\sigma^{g, N'}_R+\sigma^{s, N'}_R),
 \ee
where
$\sigma^{g, N'}_L\sigma^{g, N'}_R$ comes from the contribution of the NS-NS
part of the massless spectrum, $\sigma^{s, N'}_L\sigma^{s, N'}_R$ from the R-R
part and $\sigma^{g, N'}_L \sigma^{s, N'}_R+ \sigma^{s, N'}_R\sigma^{g, N'}_L$ 
from the NS-R and R-NS parts.

\subsection{Computation of the coefficients}\label{coeffavercomp}

We will now compute in details the decay rate (\ref{gammadom}). Special care
must be given to the issues of normalization: writing the decay rate
as (\ref{gammadom}, \ref{sigmasdom}), we must ensure that the various
contributions 
contain the correct relative normalization factors and also that the
overall normalization for the
total decay rate is correct.

This issue is solved in string theory by computing typical
tree-level amplitudes and comparing the results to well-known field
theory results. This procedure specifies the normalization of the
vertex operators and the amplitude. We will follow these rules, using
the conventions 
in \cite{Polchinski}. However, eventually, at our level of accuracy, constant
factors of order one will be neglected.

\subsubsection{Gravitons, dilatons, Kalb-Ramond fields, Kaluza-Klein,
  Winding modes and Scalars (NS-NS sector)}
\nopagebreak

In this case the relevant on-shell vertex operator $V_\zeta$ is given
by\footnote{Note also that $k^\sigma\xi_{\sigma\nu}+q^j\xi_{j\nu}=0$.}
(restoring $\alpha'$ for a moment) 
 \beq \label{gravvertex}
  V_g(k, z)= {2 \ov \alpha'} e^{i k\cdot X+i q\cdot X} 
     (\partial X^\mu-i{\alpha' \ov 2} \psi^\mu k\cdot\psi)\,
     (\bar\partial X^\nu-i{\alpha' \ov 2} \tilde\psi^\mu
  k\cdot\tilde\psi)\,\xi_{\mu\nu}, \quad k^2+q^2=0
 \eeq
which, for $\mu, \nu \in \{0, \ldots, d\}$, is (a
superposition of) the 
  graviton, dilaton, Kalb-Ramond field as well as their $D$-dimensional
  massive Kaluza-Klein or  winding versions (when there are
  compactified dimensions).

From (\ref{sigmatwoindex}), it is
found\footnote{We have formally written  
$\xi_{\mu\nu} =  \xi_{\mu} \tilde \xi_{\nu}$.} \cite{CIR} 
 \begin{align} \label{Xdecay}
  \sigma_L^{g, N'} & = \sum_{\xi} {\xi\cdot\xi (N-N')\over \sqrt{G_c(N)}}
     {1 \ov 2\pi i}\oint {dz \over z^{N'+1}}
          \,{f(z)^{2-D}\over 1-z^{N-N'}} \,
      \left( z^{-{D-2 \ov 16}} g_3(z)^{D-2}\right. \nonumber \\
         &~~~~~~~~~~~~~~~~~~~~~~~~~~~~
    \left.  - z^{-{D-2 \ov 16}} g_4(z)^{D-2}+ g_2(z)^{D-2}\right) ,
 \end{align}
where
 \begin{gather}\label{fss}
 f(z)=\prod_{n=1}^\infty (1-z^n) \\
 g_3(z)=\prod_{r={1\ov 2}}^\infty(1+z^r)\ , \ \ \
 g_4(z)=\prod_{r={1\ov 2}}^\infty(1-z^r)\, \ \ \ 
 g_2(z)=\prod_{r=0}^\infty(1+z^r) \ .
 \end{gather}
$\sigma_R$ is obtained from $\sigma_L$ substituting
left-moving quantities with right-moving ones.
The sum over the various decay products includes a sum 
over the polarizations as
 \beq
 \sum_\xi \xi\cdot \xi = D-2 \,.
 \eeq
As found in \cite{AmRus, CIR}, the loop integral over $z$ has a large
contribution when $N'\sim N$, yielding 
 \beq
  \sigma_L^{g, N'} \sim (N-N') 
    {e^{-\pi\sqrt{D-2}({N-N' \ov 2\sqrt{N}})}\over 1-e^{-\pi\sqrt{D-2}({N-N' \ov 2\sqrt{N}}) }}.
 \eeq
Expanding:
 \beq \label{estimsigmag}
  \sigma_L^g \sim {\sqrt{N} \ov \pi \,\sqrt{D-2}} \,.
 \eeq
Moreover, from
(\ref{phasespacgeneral}, \ref{emitmom}) we see that for $m=N' \sim N$ only
small values of the compact momenta and windings $\{q^i\}$ can be
excited. 
For $m \ll N$ and $p, \,\tilde p$
small, modes with higher values for the $\{q^i\}$ can
be excited, but such contributions  
are suppressed by a factor $e^{-\sqrt{N}+m}\sim e^{-\sqrt{N}}$ from 
(\ref{Xdecay}).

The contribution form the $D$-dimensional scalars and vectors (having
polarization tensor with one or both indexes in the compactified
dimensions) are subdominant, since in that case either $\sigma^{N'}_L$
or $\sigma^{N'}_R$ or both go like $\sim N-N'$, \,$N' \sim N$.

\subsubsection{Spinors and gauge fields  (R-R sector)}
\nopagebreak

Amplitudes with Ramond-Ramond states are generally difficult to compute
within string theory, and the result we present here is novel.

One of the complication we have to face concerns the choice of
pictures for the  vertex
 operators which must compensate for the superghost charge
 anomaly. Therefore, we
 must consider\footnote{In the following formulas of this section
 the index $\pm{1 \ov 2}$ for the vertex operators indicates the
 picture the operators are written in.}
 \beq \label{averagesigmaFerm}
 \sigma_L^{s, N'} = {1 \ov \sqrt{G_c(N)}}{1 \ov (2\pi i)^2}
   \sum_{s} \oint {dz \ov z^{N+1}}\oint {dz' \ov z'}z^{\prime\,N-N'}
  \text{tr}[V_{-{1 \ov 2}}(1)V_{-{1 \ov 2}}(z')z^{\hat N_L}]
 \eeq
where the on-shell physical vertex operators (left-moving part) are
 \begin{align}
 V_{-{1\ov 2}}(z, p) & = {\alpha^{\prime\, {1 \ov 4}}\ov \sqrt{2}}
       \bar u_{\dot \alpha}S^{\dot \alpha}(z)e^{-{\phi \ov 2}(z)}
      e^{i k\cdot X(z)+i q\cdot X(z)} \\
 V_{{1\ov 2}}(z, p) & = \left({2 \ov \alpha'}\right)^{{1 \ov 2}}
  \left({i\partial X^\mu(z)\ov \sqrt{2}}+{\alpha' \ov 2}{p\cdot\psi \ov 4}\psi^{\mu}\right)
   \bar u_{\dot \alpha}\gamma_{\mu\,\beta}S^{\beta}(z)
    e^{{\phi \ov 2}(z)} e^{i k\cdot X(z)+i q\cdot X(z)} \nonumber \\
   &  k^2+q^2=0\, .
 \end{align}
Here
 \begin{itemize} 
 \item $\phi$ is the bosonized ghost field 
 \item $S^{\dot \beta(\beta)}(z)$ is the {\em spin field}
 \item $u^{\dot \beta(\beta)}$ is a spinor wave function.
 \end{itemize}
The right-moving part of the vertex operator is obtained by
substituting $\bar \partial, \tilde \psi, \ldots$ to the analogous
quantities here. 

The spin field is in a representation of $SO(9, 1)$ as in
\cite{Atick:1986ns}: it 
is given by a product of spin field for $SO(2)$. Again, using a
formula such as (\ref{averagesigmaFerm}) allows us to use the same
approach as a one-loop amplitude,
of course without spin structure one loop
phases, therefore only physical states enter the
amplitude. 

The fact that the spin fields are written as tensor product of
$SO(2)$ spin fields make it straightforward to study the cases where
the number of extended dimensions $D$ is even. We will therefore limit
ourselves to that case\footnote{It would be possible to study odd extended
  dimensions, for example in the light-cone gauge,
  but at the price of decomposing their spinor fields $S^a$ in
  subgroups representation, which is an effort not worth doing for
  the present analysis.}.

Before proceeding, it is useful and interesting to observe the Lorentz
structures originating from (\ref{averagesigmaFerm}). It is
straightforward to see that the only non-zero contribution comes from
 \beq
 \text{tr}[\bar u_{\dot \alpha}S^{\dot \alpha}(1)e^{-{\phi \ov 2}(1)} e^{-ikX(1)}\partial X^\mu(z')\bar u_{\dot \alpha}\gamma_{\mu\,\beta}S^{\beta}(z') e^{{\phi \ov 2}(z')} e^{ikX(z')}]
 \eeq
whose Lorentz structure is simply
 \beq
  \bar u \slashed{P} u
 \eeq
which, after using the polarization sum\footnote{The sum over the
  various decay products (spinors with various internal indices, as
  seen form $D$ extended dimensions) gives the same result as
  summing over the ten dimensional polarizations.}
 \beq
  \sum_s \bar u^s u^s = \slashed{k}+\slashed{q}\, ,
 \eeq
leads to a factor
 \beq
 8(-\omega E+q^iQ_i)=8(N-N').
 \eeq

We therefore obtain (for our definition of the theta functions, see
the appendix \ref{thetafunc})
 \begin{align} \label{Sdecay}
  \sigma_L^{s, N'} & = {8(N-N')\over G_c(N)}{1 \ov (2\pi i)^2}
    \oint {dz \over z^{N+1}} 
    \oint {dz' \over z'^{N'+1}}\,{f(z)^{{3(2-D) \ov 2}} \ov (1-z')^{10-D \ov 8}} \,
      \left( \theta'_1(0, z) \ov 2\pi\,i\,\theta_1(z', z)\right)^{{D-2 \ov 8}} 
    \nonumber \\
   & ~~~~~~\times
   \left(\theta^{{D-2 \ov 2}}_3(\sqrt{z'}, z)-
   \theta^{{D-2 \ov 2}}_4(\sqrt{z'}, z)+
    \theta^{{D-2 \ov 2}}_2(\sqrt{z'}, z)\right) ,
 \end{align}

It is difficult at this point to give a final closed formula for this
(left-moving) quantity, since computing the loop integral over $v$ is
not straightforward in $D \neq 10$ (remember $D$ is the number of
extended dimensions). For $D=10$, in any case, using theta functions
identities, we obtain: 
 \be
  \sigma_L^{s, N'} & = & {8(N-N')\over G_c(N)}
     {1 \ov (2\pi i)^2}\oint {dz \over z^{N+1}}
       \oint {dz' \over z^{\prime\,N'+1}}\, {\theta^4_2(0, z) \ov f(z)^{12}} 
   {\theta'_1(0, z)\theta_2(z', z) \ov 2\pi\,i\,\theta_2(0, z)\theta_1(z', z)} 
     \nonumber \\
   & = &  {8(N-N')\over \sqrt{G_c(N)}}{1 \ov 2\pi i}
     \oint {dz \over z^{N'+1}}
          \,{f(z)^{8}\over 1+z^{N-N'}} \,
      g_2(z)^{8},
 \ee
where $f(w), g(w)$ has been defined in (\ref{fss}).

We see that $\sigma^s_R$ has a dominant contribution for $N' \sim N$,
as $\sigma^g_R$ had, but this time:
 \beq \label{estimsigmas}
  \sigma^{s, N'}_L \sim N-N' \, .
 \eeq
We expect similar results in any dimension $D$.

\subsubsection{NS-R and R-NS sectors}
\nopagebreak

The formula for the emission of gravitinos is given by
$\sigma^{g, N'}_R \sigma^{s, N'}_L+ \sigma^{s, N'}_L \sigma^{g, N'}_R$ with
 \be \label{estimsigmarnsnsrI}
 \sigma^{g, N'}_L && \text{given by (\ref{Xdecay})} \\
  \label{estimsigmarnsnsrII}
 \sigma^{s, N'}_L && \text{given by (\ref{Sdecay})}
 \ee
and the correspondent left-moving quantities.

\subsubsection{Total}\label{coefftotmassless}
\nopagebreak

By using (\ref{coefffromtree}, \ref{gammadom}, \ref{estimsigmag},
\ref{estimsigmas}, \ref{estimsigmarnsnsrI}, \ref{estimsigmarnsnsrII}),
 we obtain that the dominant contribution to the one-loop amplitude is
 given by

 \begin{itemize}
  \item $r=\tilde r=s=0$ 
    \,\, for the contributing coefficients\footnote{This is
    specific for the average computation. It is easy to check, by
    choosing an initial string state of preference that that is not
    generally the 
    case. Note that $r+\tilde r-s$ is related to the spin of the states.} 
  \item  
  \beq \label{domcoeff}
  \chi^{0, 0}_{-m, 0}= {2^{{3-D \ov 2}} \ov \pi}\,
     \sigma_L^{0,\, m} \approx  
   {N-m \ov e^{\pi\sqrt{D-2}{N-m \ov 2\sqrt{N}}}-1}, \qquad N-m \lesssim \sqrt{N}
  \eeq
  \item contributions with $m \ll N$ are exponentially suppressed.
 \end{itemize}

\subsection{The average mass shift} \label{avmassshift}

We are able now to sum up all that we have done so far and eventually
obtain the average closed string mass shift.
Indeed, we can correctly individuate the asymptotic behavior of the
integrals factors 
(\ref{integralnutaufin}), since we know which are the dominant
contributions. Once computed the integral asymptotically for large
$N$, we will include the coefficient factors and perform the sum over
the dominant channels in (\ref{toruscompactform}), which would give us
the leading asymptotic behavior for the amplitude.

Thanks to the analysis in the previous sections,
we have seen that the total one-loop amplitude can be subdivided in
two contributions: one corresponding to massless interactions and the
other to massive one. 
We have argued at various points (see in particular
section \ref{dominantchanneldiscuss}) that the asymptotes of these
two contributions are very different and that the first one is
dominating. 

Recall that the name ``massless interactions'' is a bit
misleading: by this we mean that at least one of the states running in
the loop is massless or a pure Kaluza-Klein/winding mode\footnote{Which are
not massless form the point of view of the extended dimensions, but
are so in ten dimensions.}.

Considering these contributions, given by
$  p=\tilde p =0 $, and taking into account section \ref{coefftotmassless},
(\ref{integralnutaufin}) reads
 \beq \label{intrealfin} 
 {1 \ov \mathcal{V}_{c}}\sum_{\{q_i\}}\text{Re}(H) \! =  \!
 {1 \ov \mathcal{V}_{c}}\sum_{\{q_i\}}
  \int_F {d^2\tau \ov \tau_2^{{d+1 \ov 2}}} \int d^2\nu 
  e^{-4\pi\left|(N+Q^2){\nu_2^2 \ov \tau_2}+(m-N)\nu_2-2 q\cdot Q\nu_2+(q^2-{n\cdot w \ov 2}) \,\tau_2  \right|}\,,
 \eeq
where we have also included the sum over winding and Kaluza-Klein
modes with the volume factor for the compactified
dimensions (in unit of $\alpha'$ and
neglecting factors of $2\pi$):
 \beq
 {1 \ov \mathcal{V}_{c}} \equiv   \prod_{i=1}^{d_c-d}\frac{1}{R_{(i)}}.
 \eeq
The absolute value in the exponent takes into account the analytical
continuation. 

It is convenient to re-write (\ref{intrealfin}) as
 \be 
 {1 \ov \mathcal{V}_{c}}\sum_{\{q_i\}}\text{Re}(H)
  \!\!\!\!& = &\!\!\!\!  {1 \ov \mathcal{V}_{c}}\sum_{\{q_i\}}
  \int_F {d^2\tau \ov \tau_2^{{d+1 \ov 2}}} \int d^2\nu 
  e^{-4\pi\left|N{\nu_2^2 \ov \tau_2}+(m-N)\nu_2+\left((q_i-Q_i{\nu_2 \ov \tau_2})^2-{n\cdot w \ov 2}\right) \tau_2\right|}\,. \nonumber \\
 \ee
From (\ref{domcoeff}) and large $\sqrt{N}= \sqrt{M_0^2-Q^2}$, the sum
over $m$ in the total amplitude (\ref{toruscompactform}) for the
dominant contribution can be approximated by an integral, using
 \beq \label{domfinalsummary}
  N-m \sim y \sqrt{N}, \qquad 0 < y < 1.
 \eeq
We obtain
 \beq \label{intmasdom}
 {1 \ov \mathcal{V}_{c}}\sum_{\{q_i\}}\text{Re}(H)  =  
 {1 \ov \mathcal{V}_{c}}\sum_{\{q_i\}}
  \int_F {d^2\tau \ov \tau_2^{{d+1 \ov 2}}} \int d^2\nu 
  e^{-4\sqrt{N}\pi\left|{\sqrt{N} \ov \tau_2}\nu_2^2-y\nu_2+\left((q_i-Q_i{\nu_2 \ov \tau_2})^2-{n\cdot w \ov 2}\right) {\tau_2 \ov \sqrt{N}}\right|}\,  .
 \eeq
In the limit $\sqrt{N} \gg 1$ the (Laplace) integral over $\nu_2$
is strongly 
dominated by $\nu_2 \sim y {\tau_2 \ov \sqrt{N}}$. Then, the integral over
$\tau_2$ favors $q_i \sim Q_i{y \ov \sqrt{N}}$ 
(especially for large $\tau_2$, which  will dominate as we will
see in a moment). Since $Q
\ll \sqrt{N}, M_0$ in the limit\footnote{Recall that $N=M_0^2-Q^2$.}
$N, M_0 \gg 1$, then the favored 
channels have $q_i \sim 0$.

We can
see that the contribution from $\tau_2 < \sqrt{N}$ has a 
different asymptotic behavior.
Indeed, recall that the integrand of the one-loop
amplitude has in effect an expansion in terms of $\sin(\pi\nu)$
starting from the power $|\sin(\pi\nu)|^{-2}$, as
dictated by the OPE in order to have the correct limit for $\nu \to
0$. 
But then, for a generic power $n$
 \beq
 |\sin(\pi\nu)|^n = 
  \sum_{l \geq 0} \left(\begin{matrix} {n \ov 2} \\ l \end{matrix}\right) 
   \sin(\pi\nu_1)^{{n\ov 2}-l}(\sinh(\pi\nu_2))^l.
 \eeq
When $\nu_2 \sim 0$ asymptotically, we see that the only term that
survives is the one where 
there are no powers of $\sinh(\pi\nu_2)$. 

Going back to the expansion in section
\ref{subsec_exponeloop}, we see that this shows up in the term with $m=N$. This
contribution to the amplitude does
not have an imaginary part, therefore its coefficient factor will have
an asymptotic behavior different from those of the terms that
contribute for $\tau_2 \geq \sqrt{N}$. In fact, 
this contribution corresponds to the excitation of a massless state with zero
momentum (in ten dimensions, 
see (\ref{phasespacgeneral}, \ref{imaginarypartform},
\ref{emitmom})). Such a term is related to a tadpole-like diagram. We
believe that it must be subdominant because of the behavior of the infrared
singularity under
supersymmetry or taking into account the Fischler-Susskind
mechanism when supersymmetry is absent\footnote{We are aware that this
  is only a supposition and not 
a rigorous proof.}. Even if this was not the case, its different
origin and
asymptotic behavior allows us a least to separate this contribution
from the fully computable remaining one.   

By rescaling $\tau_2 \to \sqrt{N} \tau_2\sqrt{1-\tau_1^2}$ 
and taking into account the comment after
formula (\ref{intmasdom}), the contribution from $\tau_2 \geq
\sqrt{N}$ gives instead 
 \be \label{masslesslargthanM0}
 {1 \ov \mathcal{V}_{c}} \sum_{\{q_i\}}\text{Re}(H) & \sim &
   {1 \ov 2\,\mathcal{V}_{c}\,\sqrt{N}} 
   \int_1^\infty {d\tau_2 \ov N^{{D -3 \ov 4}}\tau_2^{{D \ov 2}-{1 \ov 2}}}
 \int_{-{1 \ov 2}}^{{1 \ov 2}} d\tau_1 {1 \ov \sqrt{1-\tau_1}^{{D \ov 2}-{3 \ov 2}}}
    \nonumber \\
  & = &
   {c \ov (D-2)\,\mathcal{V}_{c} \,\sqrt{N}}N^{{3-D \ov 4}} \,.
 \ee
where $c$ is a (positive) 
constant (taking into account also the integration over
$\tau_1$). 

Stepping 
aside for one moment, we can see also that for contributions different from the
massless ones, so that $N-m \neq y \sqrt{N}$, the integral over
$\nu_2$ is in general asymptotically suppressed unless $\tau_2 > N$. 
The not suppressed contribution goes then like $N^{{2-D \ov 2}}$ and
this tells us again that it is subdominant to the massless
contribution given in (\ref{masslesslargthanM0}). 

\vskip 0.2cm
To obtain the total result for the real part (\ref{toruscompactform}),
we sum\footnote{We consider all the non-zero contributions and their
  relations according to (\ref{masslesslimitcoeff}). Among them, those
asymptotically dominant are the important ones.} 
over $m$ approximating, as we said, the
sum with an integral in the limit of large mass as
$ \sum_m \sim 2\sqrt{N} \int d y $ (see (\ref{domfinalsummary})
and we recall
 \beq \label{coefffinav}
 \chi \tilde \chi \sim N {y^2 \ov \left(e^{\pi{\sqrt{D-2} \ov 2} \,y}-1\right)^2}
       \, 
 \eeq
from
(\ref{domcoeff})\footnote{Written here with a bit of sloppy
  notation.}.
We
then have to double the result to take into account massless
interactions in both arcs of the loop.

We finally obtain
 \beq
   \text{Re}(T_{T^2})  \sim
      {2 \ov D-2}c'g_s^2 (M_0^2 -Q^2)^{1+{3-D \ov 4}}\,
 \eeq
where we have taken the compactified dimensions at radii $R_{(i)} \sim
\sqrt{\alpha'}=2$ and included the numerical factors in the (positive)
constant
$c'$. This result, here expressed in unit of $\alpha'$, can be rewritten in
terms of the Newton's constant in $D$ spacetime
dimensions,
which in this case is related 
to the string coupling (see\cite{Polchinski}) by 
 $G_N^D = (2\pi)^{D-3}g_s^2 (\alpha')^{{D-2 \ov 2}}$. 
We redefine now $g_s$ in order to get rid of the factor 
${2\,c' \ov D-2}$. 

Adding the result for the imaginary part from (\ref{imanddec}, 
\ref{gammadom}), see also \cite{CIR}, we obtain\footnote{We write here only
  the massless contribution to the imaginary
part.}
 \beq
  T_{T^2} = g_s^2\,(M^2-Q^2)^{1+{3-D \ov 4}}+i\pi g_s^2M^2.
 \eeq
Note that in this last formula we have expressed the mass shift and
related quantities in terms of the true mass instead than the
tree-level one, which is probably even a more accurate estimate.

We can now determine the average mass shift for closed string
states. Since we use a mostly plus metric, the field propagator for an
unstable state is\footnote{We are not interested in the details of the
  numerator.}
 \beq 
  G(p) \sim {Z \ov p^2+M^2-i\,M\Gamma},
 \eeq 
where $\Gamma$ is the total decay. Indeed the time dependence of the
propagator shows the expected exponential decay:
 \be
  G(\vec p, t) & = &\int {d p^0 \ov 2\pi} e^{-i\,p^0 t}G(p)\sim
   {Z \ov \sqrt{p^2+M^2-i\,M\Gamma}}e^{-i\sqrt{p^2+M^2-i\,M\Gamma}}
   \nonumber \\
   & \sim & {Z \ov \sqrt{p^2+M^2-i\,M\Gamma}}
      e^{-iE_{\vec p}\,t-{M \ov E_{\vec p}}\Gamma\,t}.
 \ee
Therefore, since it is $\Gamma = M^{-1} \text{Im}(T_{T^2})$, it must
be
 \beq
  p^2+M_0^2-T_{T^2}=  p^2+M_0^2-\text{Re}(T_{T^2})-i\,\text{Im}(T_{T^2})=
   p^2+M^2-i\,M\Gamma\,,
 \eeq
so that the average squared string mass shift at mass level $N$ is
 \beq
  \Delta M^2 \equiv M^2-M_0^2 = -\text{Re}(T_{T^2}) = 
   - g_s^2(M^2-Q^2)^{1+{3-D \ov 4}}.
 \eeq 
 
\section{Discussion and Conclusion}

In this work we have presented a detailed investigation
of (closed) string one-loop amplitudes in string
theory, in particular 
dealing with the average mass shift for states at a given squared mass
and Neveu-Schwarz charges.
Our algorithm is based only on well-defined string amplitudes and the
exploitation of symmetries and unitarity properties of the torus
amplitudes. 

We have been able to argue and show at various moments
(sections \ref{realpartgen}, \ref{dominantchanneldiscuss}, and after
formula (\ref{masslesslargthanM0})) that the dominant contribution to
the decay channel is given by long-range interaction (in fact gravitational
ones). 

We are aware, though, that our arguments are evidential, but not an
exhaustive mathematical proof. However they strengthen the physical
intuition that gravitational interactions
(which grows with the mass) are the dominant one for very massive states.
Furthermore, our results, coming form a direct string theory
computation, clarify and correct previous estimates of the string
self-energy, obtained so far only in approximated models which assumed
the predominance of gravitational interactions from the outset.

In the limit of large mass, we find that the average squared
mass shift for states at a given mass is
 \beq \label{masshisftconcl}
  \Delta M^2 = -g_s^2 (M^2-Q^2)^{1+{3-D \ov 4}} \,.
 \eeq 
The interaction becomes important when (consider the case $Q^2=0$)
 \beq
  g_s \sim M^{{d-6} \ov 4}\,
 \eeq
which is the expected behavior for this kind of interactions (see
formulas 3.3, 3.4 in \cite{HorPolSelf}).

Discussing (\ref{masshisftconcl}) more in details, we distinguish two
contributions to this result: the 
first one ($M^2-Q^2$) comes from the coefficients (\ref{coefffinav})
and therefore the coupling,
the second factor ($(M^2-Q^2)^{{3-D \ov 4}}$) comes from the integration over
$\tau_2$. 

Note that $\tau_2$ is linked to the range of
the interactions. Indeed if we consider the string formula
(\ref{oneloopfixedRN}) in the operator formalism (see
\cite{Green:1987mn}), it is related 
to the internal loop momentum $k$ by $e^{-\tau_2 k^2}$. We expect it,
therefore, to be related to the squared range $L^2$ of the interactions, and
looking at (\ref{intrealfin}, \ref{masshisftconcl}), we
find that this is  $L^2 \sim \sqrt{N}=\sqrt{M^2-Q^2}$. 

This result can be understood in a simple way.
We observe that a string at large mass is
well-represented at tree-level by a random-walk of size $L^2 \sim \sqrt{N}$
(\cite{chialvasize}) and  assume
a Newton potential acting
between two points on the string\footnote{The factors $L^{-1}$
  represent an averaging since it is irrelevant where along the walk the
  integrations start.} $\int {dr' \ov
  L} \int {dr\ov L} r^{D-3}$. We expect
that the integration will have an upper bound given by
the size $L$ of the string, so that the contribution is $\sim
L^{3-D}$, which is the second factor we find. 

In performing the Newton
potential calculation we have neglected the contribution that
arises from the small distances along the string (lower end of the integral),
for which a classical picture is
not suited: it is only the quantum computation
that can give us the precise result.

An interesting picture of one-loop corrections
to the string in non\! -\! supersymmetric configurations arises: the dominant
interactions responsible for the corrections are of long-range type
(namely gravitational) and it appears that
perturbations theory is generally reliable on the whole spectrum of massive
string states in any dimensions\footnote{That is to say, given a
  certain $g_s<1$, set by the dilaton vacuum expectation value, the
  correction (\ref{masshisftconcl}) is never larger than the
  tree-level value for any mass. 
  Our result does not exclude that particular sets of
  states, not representing significant portion of the string spectrum
  (sets of ``measure zero'') and therefore not affecting the average,
  could have larger corrections and therefore not be suited for a
  perturbative treatment, such as possibly the leading Regge trajectory
  states of \cite{SundbShift}.}.

\section{Acknowledgments}

I would like to thank Igor Pesando and Paolo di Vecchia for the
many conversations regarding this project. I am especially grateful to Bo
Sundborg for his comments on the results of this work.

\appendix

\section{Computation of the residue $\mathcal{R}^{r, s}_m$}\label{residuecomp}

Here we compute the residue $\mathcal{R}^{r, s}_m$ involved in the
derivation of the finite 
difference equations for the one-loop amplitude series coefficients,
see (\ref{residuedef}).
Recall its definition
 \beq \label{residueRapp}
  \mathcal{R}^{r, s}_m = 
  {1 \ov 2\pi i}\oint_{\mathcal{C}_1}{dv \ov v}v^{m+N} \mathcal{F}^{r, s}(v, w)\,,
 \eeq
where $\mathcal{C}_1$ is a circuit around $v=1$ (see figure
\ref{contours}).
In order to find $\mathcal{R}^{r, s}_m$ we need to expand
$\mathcal{F}^{r, s}(v, w)$ around $v = 1$, that is $\nu = 0$ (recall
$v\equiv e^{2\pi \, i \nu})$.

Recall the definition (\ref{defmathcalF}) of $\mathcal{F}^{r, s}(v,
w)$, here reproduced for the reader convenience
 \beq
  \left(2\pi i{\theta_{1}(\nu,\tau) \ov \theta'_{1}(0,\tau)} \right)^{2N}
  \mathcal{D}^r_s(\partial^2 \log\theta_1, \partial^3 \log\theta_1, \ldots)
  \mathcal{X}_\phi(\nu, \tau) \equiv \mathcal{F}^{r, s}(\nu, \tau)\, .
 \eeq
We know that
 \be \label{limitsmallnuthet}  
  {\theta_{1}(\nu,\tau) \ov \theta'_{1}(0,\tau)} & \sim &
    {\nu \ov 2}-{\pi^2 \ov 12}\nu^3 \\
  \label{limitsmallnuder}
  \partial_\nu^n \log\theta_1 & \sim & {a_n \ov \nu^n}\left(1+O(\nu)\right) 
 \ee
for some constant $a_n$ depending on $n$.

On the other hand, the OPE among physical vertex operators\footnote{Here we
  assume normalized 
states.} dictates the most singular term to be (left-moving part only) 
 \beq \label{OPEvetex}
  V^L_\phi(1)V^L_\phi(v) \sim {1 \ov \nu^2}
 \eeq
and since the leading singularity in the correlator is given indeed by
the OPE computation, we find
 \beq \label{limitD}
  \mathcal{D}^r_s(\partial^2 \log\theta_1, \partial^3 \log\theta_1, \ldots)
  \mathcal{X}_\phi(\nu, \tau) \sim {c \ov \nu^{2N+2-2\,r}}.
 \eeq
This is straightforward to understand by observing formulas
(\ref{PsExpansion}, \ref{PinD}, \ref{defD}) and realizing that the most
singular term in 
the expansion (\ref{PinD}) for
$\mathcal{P}(W, \Omega, \tilde\Omega,\ldots)$ comes from the term 
$\mathcal{D}^0_0(\partial^2 \log\theta_1, \partial^3 \log\theta_1,\ldots)
\mathcal{D}^0_0(\bar\partial^2 \log\bar\theta_1, \bar\partial^3
\log\bar\theta_1, \ldots)$ because of
(\ref{limitsmallnuder}). Terms with $r, \tilde r \neq 0$ will have less
singular behavior for $\nu \to 0$ because they contain $r$ double
derivative of $\log\theta_1$ {\em less} than the case $r=\tilde r=0$
according to (\ref{PsExpansion}, \ref{defD}), 

Therefore, from (\ref{residueRapp}, \ref{limitsmallnuthet},
\ref{limitD}) we obtain 
 \be
  \mathcal{R}^{0, 0}_m & = & 
  {1 \ov 2\pi i}\oint_{\mathcal{C}_1}{dv \ov v}v^{m+N} \mathcal{F}^{0, 0}(v, w) \\
  & = &
  {1 \ov 2\pi i}\oint{d\nu \ov \nu}
    e^{2\pi(m+N)\, i\nu} {A^0 \nu^{2N} \ov \nu^{2N+2}}Z_{T^2} = 0. \nonumber
 \ee
for a supersymmetric background ($Z_{T^2}$ is the relevant partition
function). Note that in a non -supersymetric case we would have found
$\mathcal{R}^{0, 0}_m=A^0(m+N)$, where $A^0$ is a constant that for
normalized states is 
$A^0=1$ because 
of the OPE (\ref{OPEvetex}).

Similarly we find
 \be
  \mathcal{R}^{r\neq 0, s}_m & = & 
  {1 \ov 2\pi i}\oint_{\mathcal{C}_1}{dv \ov v}v^{m+N} \mathcal{F}^{r, s}(v, w) \\
  & = &
  {1 \ov 2\pi i}\oint{d\nu \ov \nu}
    e^{2\pi(m+N)\, i\nu} {A^{r, s} \nu^{2N} \ov \nu^{2N+2-2r}} = 0,
    \quad \forall r\geq 1. \nonumber
 \ee
We have then obtained (\ref{residueRresults}).

The fact that poles at $v=1$ are not present in supersymmetric
backgrounds also clarifies the results in
(\ref{masslesslimitcoeff}). Indeed, the fact that massless channels
with $m < -2N,\, p=\tilde p=0$ do not contribute to the one-loop
amplitude, having a zero coefficient factor could seem at first sight
perplexing. But it is simple to understand it once one realizes that the
amplitude has an expansion in powers of $(1-v)$ and  
only positive powers can occur, in order not to have poles
at $v=1$. By looking at (\ref{oneloopfixedRN}), it is possible to see
that the maximum power to appear in the limit $w \to
0$ is $(1-v)^{2(N-1)}$. 

The same should occur to massless contributions having $m=p \neq 0,\,
\tilde p=p+\sum_i n_iw_i $. We can verify it by
using the formulas for the theta functions in appendix
\ref{thetafunc}. By changing variables to 
$x_1=e^{2\pi i(\tau-\nu)},\,x_2=e^{2\pi i\nu}$, it is straightforward to see that
those contributions are accounted for by the term
of order
$x_2^0$ in (\ref{oneloopfixedRN}) which
can be written as a series expansion in $1-w$
with no negative powers and maximum power $(1-w)^{2(N-1)}$. 

\section{Arguments for suppression due to averaging}\label{eurargkin} 

We show here how simple
arguments can make us understand why averaging 
cannot individuate a preferred contribution channel to the
one-loop coefficients. We concentrate on those contributions to the
amplitude which have a non-zero imaginary part. In those cases the
coefficient of expansion (\ref{toruscompactform}) are related to the
decay rates as in (\ref{coefffromtree}). Let us concentrate on the
latter, then.

Recall
 \beq \label{gendecayaverageappendix}
 \Gamma_i = g_s^2 \mathcal{P}
    \sigma_L \times \sigma_R\, ,
 \eeq 
which has been defined in section \ref{dominantchanneldiscuss} and by
formula (\ref{sigmatwoindex}) which we re-write here for the reader's
convenience:
  \be \label{sigmatwoindexapp} 
  \sigma^{\aleph, N'}_{L} =
   {1 \ov \sqrt{G_c(N)}}
  \sum_{\zeta_{|_\aleph}} \oint {dz \ov z^{N+1}}\oint {dz' \ov z'}z^{\prime\,N-N'}
      \,\text{tr}[V^\dag_{\zeta,\,L}(1)V_{\zeta,\,L}(z')\,z^{\hat N_L}]. 
  \nonumber  
 \ee
As we said in section \ref{dominantchanneldiscuss}, apparently the
suppressing factor $G_c(N)^{-1} \sim e^{-2\pi\sqrt{d-1}\sqrt{N}}$ in
(\ref{gendecayaverageappendix}) could lead to a kinetic 
suppression of many contributions.

In order to investigate when this is the case, let us further study
the quantity  $\sigma_{R, (L)}$.
It is straightforward to realize that one gets
 \be
  \text{tr}[V^\dag_{\zeta,\,R}(1)V_{\zeta,\,R}(v)\,w^{\hat N}] & = &
     f(w)^{2-D}F(v, w) \\
  f(w) & = & \prod_{n=1}^\infty \,(1-w^n) \nonumber 
 \ee 
where the factor $f(w)^{2-D}$ comes from the contractions among the
(bosonic part of the)
states running in the trace (alternatively one reaches the same
conclusion by realizing that (\ref{sigmatwoindexapp}) is a one-loop
amplitude projected down to some definite mass levels and without
interaction over zero modes of the fields). This factor is
where one expects the large degeneracies to come from. 

$F(v, w)$ is a function of $\Omega$ defined in
(\ref{OmegaW}) and its higher 
derivatives (and {\em not} of $W$ as instead $P$ in
(\ref{oneloopfixedRN})). 

By looking at the definition of $\Omega$ (and therefore of its derivatives), we
see that the term of order $v^{N'-N}$ picked up by 
the $v$ loop integration in (\ref{sigmatwoindex}) is accompanied
(generally) by a factor $w^{(N-N')\ell}, \,\, \ell \geq 1$. 
Therefore the final loop integral
becomes $\sim \oint dw \, w^{N(\ell-1)-N'-1}$ and we see that we obtain
the largest contribution $\sim e^{\pi\sqrt{d-1}\sqrt{N'}}$ for $\ell=1$. 

On the other end we have to sum
over all possible $|\zeta\rangle$ at mass level $\aleph\sim N-N'$ with
degeneracy 
$e^{\pi\sqrt{d-1}\sqrt{N-N'}}$and this implies a 
total result of $e^{\pi\sqrt{d-1}(\sqrt{N-N'}+\sqrt{N'})}$. Similar arguments apply
to $\sigma_L$. The product of these two results compensates the
suppressing factor $G_c(N)^{-1} \sim e^{-2\pi\sqrt{d-1}\sqrt{N}}$ in
(\ref{gendecayaverageappendix}).

\section{Theta functions}\label{thetafunc}

Our conventions for the theta functions are
 \beq
   \theta_1  =  i\,
   \sum_{n=0}^{\infty}(-1)^n
     w^{{(n + {1 \ov 2})^2 \ov 2}}\,(v^{n + {1 \ov 2}} -v^{-n - {1 \ov 2}}) 
   \qquad \theta_2  =    \sum_{n=0}^{\infty}
     w^{{(n + {1 \ov 2})^2 \ov 2}}\,(v^{n + {1 \ov 2}} +v^{-n - {1 \ov 2}}) 
  \nonumber
 \eeq
 \beq
   \theta_3  =  1 + \sum_{n=1}^{\infty} w^{{n^2 \ov 2}}\,(v^{n}+v^{-n}) 
   \qquad
   \theta_4  =  1 + \sum_{n=1}^{\infty}(-1)^n w^{{n^2 \ov 2}}\,(v^{n}+v^{-n}). 
  \nonumber
 \eeq



\end{document}